\documentclass[10pt]{article}
\pdfoutput=1

\usepackage[usenames,dvipsnames,table]{xcolor}
\usepackage[utf8]{inputenc}
\usepackage{booktabs}
\usepackage[mathscr]{euscript}

\usepackage{jheppub} 

\usepackage{tikz-cd} 
\usepackage{circuitikz}
\usepgfmodule{shapes}
\usetikzlibrary{decorations.pathmorphing}
\usetikzlibrary{decorations.pathreplacing,decorations.markings,snakes}
\usetikzlibrary{backgrounds, arrows,calc,shapes,decorations.pathreplacing, automata,positioning,fit,calc}
\usetikzlibrary{calc,decorations.pathreplacing,decorations.markings}

\usepackage{cancel} 
\usepackage{xcolor}
\usepackage{enumitem}

\usetikzlibrary{shapes.misc}

\tikzset{cross/.style={cross out, draw=black, minimum size=5*(#1-\pgflinewidth), inner sep=0pt, outer sep=0pt},
cross/.default={2pt}}

\tikzset{snake it/.style={decorate, decoration=snake}}

\tikzset{
  hyper/.style    = { thick, double, 
  double distance = 3pt }}

\pgfdeclarepatternformonly{coarsedots}{\pgfqpoint{-1pt}{-1pt}}{\pgfqpoint{5pt}{5pt}}{\pgfqpoint{12pt}{12pt}}{
    \pgfpathcircle{\pgfqpoint{.5pt}{.5pt}}{.5pt}
    \pgfusepath{fill}}
    
\tikzset{gauge/.style={rounded rectangle, draw=black!100, thick, minimum size=5mm},  gaugeD/.style={rounded rectangle, draw=black!100,double,thick,minimum size=5mm},  empty/.style={rounded rectangle, draw=white!100, thick, minimum size=5mm}, flavor/.style={rectangle, draw=black!100, thick, minimum size=5mm},flavorD/.style={rectangle, draw=black!100, double,thick, minimum size=5mm}}

\definecolor{cobalt}{rgb}{0.0, 0.28, 0.67}
\definecolor{penred}{RGB}{200,20,20}
\definecolor{penblue}{RGB}{20,60,160}

\usepackage{graphicx}
\usepackage{amsmath, amssymb}
\usepackage{youngtab}
\usepackage{changepage}

\newcommand{\be}{\begin{eqnarray}}
\newcommand{\ee}{\end{eqnarray}}
\newcommand{\ba}{\begin{array}}
\newcommand{\ea}{\end{array}}
\newcommand{\bea}{\begin{eqnarray}}
\newcommand{\eea}{\end{eqnarray}}
\newcommand{\bpic}{\begin{tikzpicture}}
\newcommand{\epic}{\end{tikzpicture}}

\newcommand{\nn}{\nonumber}
\newcommand{\bn}{\begin{enumerate}}
\newcommand{\en}{\end{enumerate}}

\def\CC{\mathscr{C}}
\def\VV{\mathscr{V}}
\def\FF{\mathscr{F}}
\def\SS{\mathscr{S}}

\title{Mirror symmetry on a circle}

\author[a]{Siqi Chen,}
\author[a]{Shehab Hossam Fadda,}
\author[b]{Matteo Sacchi}

\affiliation[a]{C.N.~Yang Institute for Theoretical Physics, Stony Brook University, Stony Brook, NY 11794, USA}
\affiliation[b]{Simons Center for Geometry and Physics, Stony Brook University, Stony Brook, NY 11794, USA}

\emailAdd{siqi.chen.1@stonybrook.edu}
\emailAdd{shihab.fadda@stonybrook.edu}
\emailAdd{msacchi@scgp.stonybrook.edu}

\abstract{We investigate the small circle, or high temperature, limit of the supersymmetric index identities for the three-dimensional abelian mirror symmetry of SQED. There exist two possible limits, depending on how the parameters of the theory are scaled with the radius of the circle. In both cases the result is qualitatively similar. One side always reduces to the sphere partition function of a two-dimensional $\mathcal{N}=(2,2)$ gauged linear sigma model (GLSM). The opposite side has a two-fold interpretation, either as the sphere partition function of the Landau--Ginzburg (LG) model that is Hori--Vafa dual to the GLSM, or as a Coulomb gas integral for a correlation function of Liouville or Toda CFT. This approach thus provides a systematic way to generate integral identities between partition functions of GLSMs on the one hand, and partition functions of LG models or CFT Coulomb gas integrals on the other. The latter perspective finds useful applications in the recently proposed 2d/2d correspondence, which relates sphere partition functions of unitary 2d $\mathcal{N}=(2,2)$ theories and correlation functions of non-unitary 2d CFTs that both descend from compactifications of a unitary 4d $\mathcal{N}=2$ SCFT. We present an example based on the $(A_{k-1},A_{N-1})$ Argyres--Douglas theories, where the CFT is a non-unitary minimal model. We also give a purely two-dimensional derivation of the identities obtained in the small circle limit which is inspired by the Kapustin--Strassler piecewise derivation of 3d abelian mirror symmetry.
}

\begin{document} 

\maketitle

\flushbottom


\section{Introduction and outlook}

Infrared (IR) dualities are one of the most intriguing phenomena that can characterize quantum field theories (QFTs). They encode a redundancy in the microscopic description of the low energy physics, in that they relate two different theories which turn out to flow to the same fixed point at long distances. One of the prime examples is Seiberg duality for the four-dimensional quantum chromodynamics with minimal supersymmetry (SQCD) \cite{Seiberg:1994pq}. Dualities are crucial for our understanding of quantum field theory, since they can be used to recast a difficult question in one theory into a more tractable one in the dual frame.

Despite being unexpected at first sight, we know by now a plethora of examples of infrared dualities, especially with supersymmetry. A great deal of effort has gone into trying to find organizing principles behind these examples. There are two important approaches to this question, which will be relevant for us. Both of them rely on the idea that there is a restricted set of more fundamental dualities from which all the others can be derived. However, in the first approach we consider a renormalization group flow across dimensions to derive a lower dimensional duality from reduction of a higher dimensional one. Instead, in the second approach we work in a fixed number of dimensions and apply iteratively a basic duality to derive a new one.

Another related but distinct concept is that of a correspondence. This still relates different quantum systems, however only certain specific observables match between the two sides and there is no claim that the entire physics captured by the two theories is the same. Such theories might even live in a different number of dimensions. One notable example is the AGT correspondence \cite{Alday:2009aq,Wyllard:2009hg}, relating the four-sphere partition function of certain four-dimensional $\mathcal{N}=2$ superconformal field theories (SCFTs) known as class $\SS$ theories \cite{Gaiotto:2009hg,Gaiotto:2009we} and correlation functions of Toda conformal field theory. 

One key feature of many correspondences is that changing observable on one side completely changes the theory on the other. In the case of AGT, choosing a different Toda correlator alters the class $\SS$ theory whose $S^4$ partition function computes it, while considering the superconformal index of the class $\SS$ theory instead of its $S^4$ partition function leads to a topological quantum field theory (TQFT) \cite{Gadde:2009kb,Gadde:2011ik,Gadde:2011uv}. This fact is often understood in terms of a common higher dimensional origin of the two theories related by the correspondence, where varying the observable of one theory amounts to changing the type of compactification of the higher dimensional theory that we are performing.

Even though infrared dualities and correspondences are very different in nature, in some cases they can actually be related and coexist. In this work, we will explore one such occurrence.

Our focus will be on field theories in three and two dimensions. Three-dimensional supersymmetric dualities can be of two main types: Seiberg-like dualities and mirror symmetry. The former owe their name to the fact that they share most of their properties with the four-dimensional Seiberg dualities. In fact most of the known 3d Seiberg-like dualities, such as Aharony duality \cite{Aharony:1997gp}, can be derived from circle compactification of 4d Seiberg dualities \cite{Aharony:2013dha,Aharony:2013kma}. Mirror symmetry \cite{Intriligator:1996ex,Hanany:1996ie,Aharony:1997bx,deBoer:1997kr} instead exhibits features that are inherent to three-dimensional physics, even though recently it was shown that a 4d counterpart exists from which one can derive 3d mirror symmetry by circle reduction \cite{Hwang:2020wpd}.

The original setting for mirror symmetry was that of 3d $\mathcal{N}=4$ theories. Such theories have a moduli space of vacua that consists of two main branches, known as the Higgs and the Coulomb branch, which are both Hyperkähler. Accordingly, they have an $SU(2)_H\times SU(2)_C$ R-symmetry, where each factor is naturally associated with one of the two branches, in the sense that these two spaces are equipped with an $SU(2)_H$ or $SU(2)_C$ triplet of moment maps, respectively. They also have two different types of deformation parameters, mass parameters and Fayet–Iliopoulos (FI), which provide resolutions of the Higgs and Coulomb branch singularities, respectively. Mirror symmetry acts by swapping each such pair of objects between the dual theories
\begin{align}\label{eq:mapMS}
    \text{Higgs}\,\,&\longleftrightarrow\,\,\text{Coulomb}\nn\\
    SU(2)_H\,\,&\longleftrightarrow\,\,SU(2)_C\nn\\
    \text{masses}\,\,&\longleftrightarrow\,\,\text{FIs}\,.
\end{align}
One of the original perspectives on the power of mirror symmetry was that it maps the Coulomb branch of one theory, which receives quantum corrections and is described by disorder monopole operators, to the Higgs branch of the dual, which is instead protected by supersymmetry against quantum corrections and is parametrized by composite operators of the elementary fields.

One can get a better handle on supersymmetric dualities by considering various kinds of partition functions which can be computed exactly with supersymmetric localization, see e.g.~\cite{Pestun:2016zxk} for a review and references therein. A duality implies a non-trivial integral identity between the partition functions of the dual theories. Moreover, such partition functions typically depend on the deformation parameters of the theory and can thus be used to study how they are mapped across a duality. The advent of supersymmetric localization has thus sparked a rich interplay between the study of IR dualities in physics and that of special functions and their integral identities in mathematics, see e.g.~\cite{2003math......9252R,Spiridonov:2009za,Spiridonov:2011hf,Krattenthaler:2011da}. In this work we will see some new examples of this connection.

In three dimensions, one observable that can be computed exactly is the superconformal index \cite{Bhattacharya:2008zy,Kim:2009wb,Imamura:2011su,Kapustin:2011jm,Dimofte:2011py}, which can be viewed as a refinement of the Witten index. It also coincides with the partition function of the theory on $S^2\times S^1$ when evaluated with the superconformal R-symmetry. The latter, however, can be computed even with an R-charge assignment which is not the superconformal one. In such a case, we will refer to the resulting object as the supersymmetric index.

The supersymmetric index encodes information about the global symmetries of the theory via fugacities, i.e.~holonomies for the background gauge fields along the $S^1$, and background magnetic fluxes through the $S^2$. Mirror symmetry then implies a non-trivial identity for the indices of the dual theories, in which such parameters are identified between the two sides according to the general mapping \eqref{eq:mapMS}, where mass parameters are encoded in background fields for the flavor symmetries while FIs in background fields for the topological or magnetic symmetries.\footnote{We recall that in $d$ spacetime dimensions, for each $U(1)$ factor in the gauge group there is a $(d-3)$-form $U(1)$ magnetic symmetry with current $\star F$ \cite{Gaiotto:2014kfa}. In $d=3$ this is a 0-form symmetry, sometimes also called topological symmetry.}

We will be interested in the limit of the index corresponding to shrinking the circle to zero size $\beta\to0$, where $\beta$ is the ratio of the radii of $S^1$ and $S^2$ 
\begin{equation}
    \beta\to0\,,\qquad \beta=\frac{R_{S^1}}{R_{S^2}}\,.
\end{equation}
This limit is also known as the high temperature or Cardy limit of the index, and it has received a lot of attention, especially in the context of AdS${}_4$/CFT${}_3$ holographic microstate counting (see e.g.~\cite{Choi:2019zpz,Bobev:2019zmz,Benini:2019dyp,Nian:2019pxj,Choi:2019dfu,GonzalezLezcano:2022hcf,Bobev:2022wem,BenettiGenolini:2023rkq,Amariti:2023ygn,Bobev:2024mqw,ArabiArdehali:2025bub}). When taking such a limit for the index of a single 3d theory, the result typically diverges and the main interest when studying the Cardy limit is usually in the leading behaviour of this divergence as well as corrections. However here we will be concerned with taking the small circle limit of the index identity implied by a 3d duality. In this setup, when considering the leading divergent contribution, the entire $\beta$-dependence is an identical overall prefactor on both sides, which can thus be simplified resulting in a different and finite integral identity.

This is the perspective that has been taken for example in \cite{Aharony:2017adm,Nedelin:2017nsb,Pasquetti:2019uop}, where it was shown that starting from a 3d gauge theory the result of the limit of its index is qualitatively different depending on how the masses and FI parameters of the theory are scaled with $\beta$. There are two main types of scalings that one can consider, which roughly behave as follows:
\begin{itemize}
    \item \textbf{Limit 1}: The masses $M$ are kept finite while the FIs $\eta$ scale as $\beta$
    \begin{equation}
        M\, \,\text{ fixed}\,,\qquad \eta\sim\beta \tau\,\text{ with }\, \tau\, \,\text{ finite}\,.
    \end{equation}
    In this limit the 3d index reduces to the two-sphere partition function \cite{Gomis:2012wy} of a 2d $\mathcal{N}=(2,2)$ Landau--Ginzburg (LG) model, as shown in \cite{Aharony:2017adm}. Hence, such a LG model can be viewed as the circle reduction of the original 3d theory when such a scaling of parameters is considered. The result can also be interpreted as the correlation function of a 2d CFT in the Coulomb gas or free field representation, typically of Toda or Liouville CFT \cite{Nedelin:2017nsb,Pasquetti:2019uop}. This latter observation is closely related to the works \cite{Nieri:2013yra,Aganagic:2013tta,Aganagic:2014oia}, where it was observed that the $D_2\times S^1$ partition functions of 3d gauge theories, also known as holomorphic blocks \cite{Pasquetti:2011fj,Beem:2012mb}, can take the form of Coulomb gas integrals for conformal blocks of $q$-deformed Toda or Liouville CFT.
    \item \textbf{Limit 2}: The masses scale as $\beta$ while the FIs are kept finite
    \begin{equation}
        M\sim\beta m\,\,\text{ with }\,\,m\,\,\text{ finite}\,,\qquad \eta\, \,\text{ fixed}\,.
    \end{equation}
    Again in \cite{Aharony:2017adm} it was shown how the result can be interpreted as the $S^2$ partition function \cite{Benini:2012ui,Doroud:2012xw} of a 2d $\mathcal{N}=(2,2)$ gauged linear sigma model (GLSM) that is identical to the 3d gauge theory we started with. We thus interpret this GLSM as the circle reduction of the original 3d theory, but with this different scaling of parameters. 
\end{itemize}

As pointed out in \cite{Nedelin:2017nsb,Pasquetti:2019uop}, given the identity between the indices of 3d dual theories, the nature of the resulting identity in the $\beta\to0$ limit will be different depending on the type of duality we started with. In a Seiberg-like duality, unlike mirror symmetry, masses and FI parameters are not swapped. Hence, we will get the same type of behavior on both sides. For example, Limit 1 will result in the identity between the $S^2$ partition functions of the GLSMs arising from circle reduction of the dual 3d gauge theories. This can be interpreted as the manifestation of a 2d IR duality. However, as stressed in \cite{Aharony:2016jki,Aharony:2017adm}, claiming a duality in two dimensions is subtle when the target space is non-compact. The more conservative statement is then that the result should be viewed as a duality between mass deformed 2d theories, which is the regime that the partition function is really probing. Conversely, Limit 2 yields an identity between the $S^2$ partition functions of two LG models, or between CFT Coulomb gas integrals. This is the perspective that was taken in \cite{Pasquetti:2019uop}, where it was shown how known 3d dualities reduce in the $\beta\to0$ limit to integral identities used to study Coulomb gas integrals for Liouville or Toda CFT \cite{Goulian:1990qr,Baseilhac:1998eq,Fateev:2007qn,Fateev:2007ab,Fateev:2008bm}. Surprisingly, it is sometimes possible to reverse this logic to deduce new 3d dualities starting from known Coulomb gas integral identities \cite{Pasquetti:2019tix}, which can even be further uplifted to 4d dualities (see e.g.~\cite{Bottini:2022vpy}). 

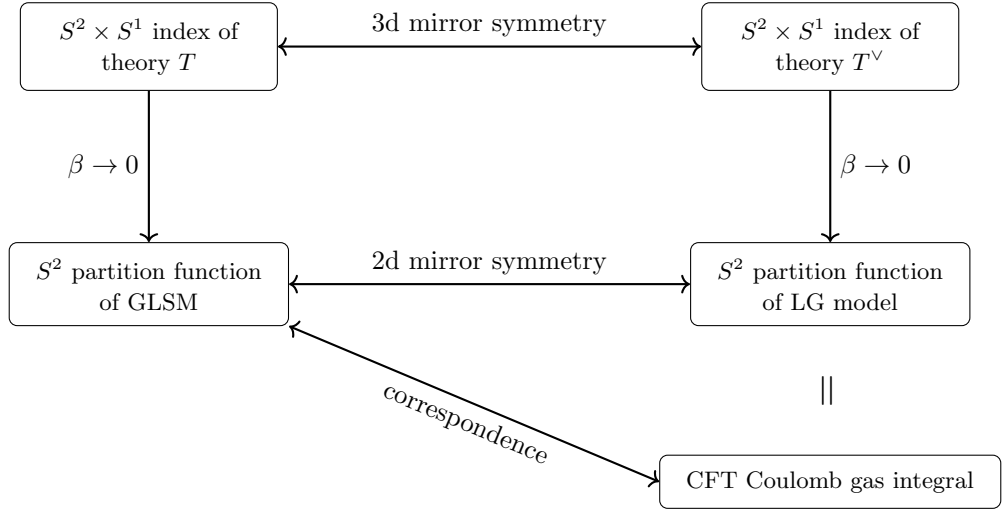
\begin{figure}[t]
\centering
\begin{tikzpicture}[
  node distance=2.0cm and 5.6cm,
  box/.style={
    draw,
    rounded corners=3pt,
    align=center,
    inner xsep=10pt,
    inner ysep=6pt,
    font=\small,
    minimum width=3.4cm
  }
]

\node[box] (T) {$S^2\times S^1$ index of \\theory $T$};
\node[box, right=of T] (Tv) {$S^2\times S^1$ index of \\theory $T^\vee$};

\draw[<->, thick] (T) -- node[above] {3d mirror symmetry} (Tv);

\node[box, below=of T] (GLSM) {$S^2$ partition function\\of GLSM};
\node[box, below=of Tv] (LG) {$S^2$ partition function\\of LG model};

\draw[->, thick] (T) -- node[left] {$\beta\to0$} (GLSM);
\draw[->, thick] (Tv) -- node[right] {$\beta\to0$} (LG);

\draw[<->, thick] (GLSM) -- node[above] {2d mirror symmetry} (LG);

\node[box, below=1.7cm of LG] (CG) {CFT Coulomb gas integral};

\node[rotate=90] at ($(LG.south)!0.5!(CG.north)$) {\LARGE$=$};

\draw[<->, thick] (GLSM.south east) -- node[below, sloped] {correspondence} (CG.west);

\end{tikzpicture}
\caption{Schematic summary of the small circle limit of the $S^2\times S^1$ index identity for a 3d mirror pair. The equality of indices between $T$ and $T^\vee$ reduces, for both possible scalings of parameters of Limit 1 and Limit 2, to an identity between the $S^2$ partition function of a GLSM and that of its mirror LG
model. Equivalently, the LG partition function can be interpreted as a Coulomb gas integral for a two-dimensional CFT, yielding a correspondence between GLSM and CFT.}
\label{fig:diagram}
\end{figure}

Our interest will be instead in the small circle limit of the index identity for a 3d mirror pair. Due to the mirror map \eqref{eq:mapMS} which swaps masses and FI parameters, both Limit 1 and Limit 2 will lead to an identity between the $S^2$ partition function of a GLSM on one side, and that of a LG model or equivalently a CFT Coulomb gas integral on the other. Depending on which of the two perspectives we choose, the result can be interpreted in two different ways, as summarized in Figure \ref{fig:diagram}.

On the one hand, we can think of it as a realization at the partition function level of 2d mirror symmetry \cite{Hori:2000kt}, which relates a GLSM to a dual LG model. This shows how 2d mirror symmetry can be obtained from circle reduction of 3d mirror symmetry \cite{Aharony:2017adm}. On the other hand, we can regard it as the equality between the partition function of a GLSM and a CFT Coulomb gas integral, which is instead a relation that is more characteristic of a correspondence rather than a duality. Indeed, changing the seed 3d duality modifies the resulting GLSM on one side and the correlation function on the other, however the latter could still be for the same 2d CFT. As we shall see, this kind of correspondence is characterized by the following dictionary between the parameters of the GLSM and of the vertex operators of Liouville or Toda CFT:
\begin{align}
    \text{FI}\,\,&\longleftrightarrow\,\,\text{position}\nn\\
    \text{masses}\,\,&\longleftrightarrow\,\,\text{momenta}\,.
\end{align}

Examples of such a correspondence have already been studied in various works \cite{Doroud:2012xw,Gomis:2014eya,Gomis:2016ljm,Nedelin:2017nsb}. In particular, both the 2d mirror symmetry interpretation and the GLSM/CFT correspondence interpretation of the small circle limit of mirror symmetry have been investigated in \cite{Nedelin:2017nsb} for the case of the 3d $\mathcal{N}=4$ $T[SU(N)]$ theory \cite{Gaiotto:2008ak}, but at the level of the hemisphere partition functions \cite{Honda:2013uca,Hori:2013ika} and conformal blocks. Here we shall instead be interested in an abelian mirror pair, and the resulting identities at the level of the $S^2$ partition functions and full CFT correlators.

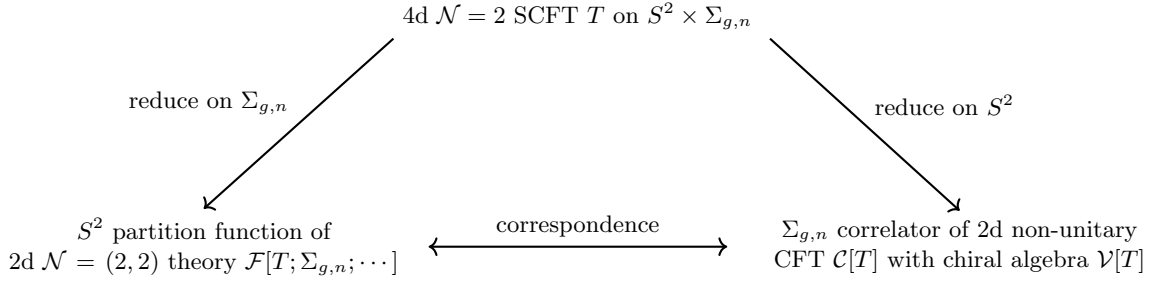
\begin{figure}[t]
\centering

\newcommand{\diagramwidth}{5}

\begin{tikzpicture}[
  every node/.style={align=center},
  font=\small
]

\pgfmathsetlengthmacro{\leftboxwidth}{1.14*\diagramwidth cm}
\pgfmathsetlengthmacro{\rightboxwidth}{1.18*\diagramwidth cm}

\node (fourD) at (0,1.7) {
  4d $\mathcal N=2$ SCFT $T$ on $S^2\times\Sigma_{g,n}$
};

\node[text width=\leftboxwidth] (classF) at (-\diagramwidth,-1.35) {
  $S^2$ partition function of\\
  2d $\mathcal{N}=(2,2)$ theory
  $\mathcal F[T;\Sigma_{g,n};\cdots]$
};

\node[text width=\rightboxwidth] (cft) at (\diagramwidth,-1.35) {
  $\Sigma_{g,n}$ correlator of
  2d non-unitary \\ CFT $\mathcal C[T]$
  with chiral algebra $\mathcal V[T]$
};

\draw[->, thick] (fourD.south west) -- node[above left] {reduce on $\Sigma_{g,n}$} (classF.north);
\draw[->, thick] (fourD.south east) -- node[above right] {reduce on $S^2$} (cft.north);

\draw[<->, thick] (classF.east) -- node[above] {correspondence} (cft.west);

\end{tikzpicture}

\caption{Higher-dimensional origin of the 2d/2d correspondence. Starting from a parent
$4$d $\mathcal N=2$ SCFT $T$ on $S^2\times\Sigma_{g,n}$, reduction on
$\Sigma_{g,n}$ gives a $2$d $\mathcal N=(2,2)$ theory whose $S^2$ partition
function is related to a correlator on $\Sigma_{g,n}$ of a non-unitary $2$d CFT
obtained by reduction on $S^2$.}
\label{fig:diagram2d2d}
\end{figure}

The second perspective turns out to be closely connected to the recently proposed 2d/2d correspondence \cite{Rastelli:2025nyn}. This states that the $S^2$ partition function of a family of 2d $\mathcal{N}=(2,2)$ unitary SCFTs labelled by a Riemann surface $\Sigma_{g,n}$ of genus $g$ and with $n$ punctures, dubbed class $\FF$ theories, can be computed by correlation functions on $\Sigma_{g,n}$ of primary operators $O_i$ in non-unitary 2d CFTs\footnote{The $\sim$ symbol is due to an ambiguity in the definition of the $S^2$ partition function of (2,2) theories \cite{Jockers:2012dk,Gomis:2012wy,Gerchkovitz:2014gta}.}
\begin{equation}\label{eq:2d2didintro}
    \mathcal{Z}_{S^2}[\FF[T;\Sigma_{g,n};O_1,\cdots,O_n]]\sim\langle O_1\cdots O_n\rangle_{\Sigma_{g,n}}^{\CC[T]}\,.
\end{equation}
This correspondence can be derived from a higher dimensional perspective, as summarized in Figure \ref{fig:diagram2d2d}. Starting from a parent 4d $\mathcal{N}=2$ SCFT $T$ on a certain $S^2\times \Sigma_{g,n}$ background, one can consider two possible limits. On the one hand, we can reduce on $\Sigma_{g,n}$ to get the 2d $\mathcal{N}=(2,2)$ class $\FF$ theory on $S^2$. The specific theory will depend on the choice of 4d SCFT $T$, of the Riemann surface $\Sigma_{g,n}$, and of additional data that should be specified at the punctures, which are related to the choice of primaries $O_i$ in the non-unitary CFT. For this reason, we denote these theories by $\FF[T;\Sigma_{g,n};O_1,\cdots,O_n]$. On the other hand, we can reduce on $S^2$ to get the 2d CFT on $\Sigma_{g,n}$. In this case, choosing a different $T$ will change the 2d CFT and so we denote it by $\CC[T]$, while varying $\Sigma_{g,n}$ will modify which correlator we are computing. We see again that different choices of the compactification surface $\Sigma_{g,n}$ lead to different theories on the unitary side, while on the non-unitary side we still have the same theory but we are computing a different observable.

This 2d/2d correspondence can be regarded as a lower dimensional analog of the AGT correspondence, with the class $\FF$ theories being a two-dimensional version of the four-dimensional class $\SS$ theories. It also provides an upgrade of the SCFT/VOA correspondence of \cite{Beem:2013sza}. The latter associates to the 4d $\mathcal{N}=2$ SCFT $T$ a chiral algebra $\VV[T]$. The 2d non-unitary CFT $\CC[T]$ has as its chiral algebra exactly the one $\VV[T]$ involved in the SCFT/VOA correspondence. It thus promotes the chiral algebra to a full-fledged 2d CFT. 

In many cases, the CFT correlation functions involved in the 2d/2d correspondence admit a Coulomb gas integral representation. One can then obtain a GLSM description of the class $\FF$ theories by first finding a suitable 3d gauge theory whose index reduces in the $\beta\to0$ limit to the Coulomb gas integral at hand, and then considering the limit of the index of its mirror. In \cite{Rastelli:2025nyn} this strategy was used for the family of 4d $\mathcal{N}=2$ Argyres--Douglas SCFTs $T=(A_1,A_{2k})$ \cite{Argyres:1995jj,Argyres:1995xn,Xie:2012hs}, and with $\Sigma_{0,n}$ a sphere with $n$ punctures. Since the CFTs associated with these Argyres--Douglas theories are Virasoro minimal models, we can view their correlation functions as special cases of the Liouville correlators when the momenta of the vertex operators take rational values and we can thus find Coulomb gas integral representations for them. 

This illustrates how, in this particular setting, dualities and correspondences become intimately intertwined. One of the goals of the present work is to systematically explore the identities among GLSM partition functions, LG partition functions and Coulomb gas integrals that arise from the small circle limit of 3d mirror symmetry, and to show how to apply them in the context of the 2d/2d correspondence.

\subsection*{Summary of results}

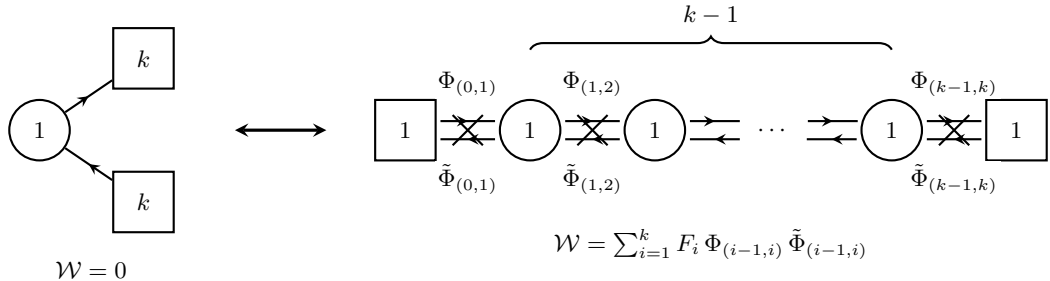
\begin{figure}
    \centering
    \begin{tikzpicture}[
    x=1cm,y=1cm,
    >=stealth,
    thick,
    every node/.style={font=\small},
    gauge/.style={draw,circle,minimum size=8mm,inner sep=0pt},
    flavor/.style={draw,rectangle,minimum size=8mm,inner sep=0pt},
    midarrow/.style={
      postaction={decorate},
      decoration={markings, mark=at position 0.5 with {\arrow{>}}}
    },
    phiLabel/.style={
      midway,
      above=\fieldlabelsep,
      fill=white,
      inner sep=1pt
    },
    tphiLabel/.style={
      midway,
      below=\fieldlabelsep,
      fill=white,
      inner sep=1pt
    }
]

\def\xunit{1.1}
\def\crosssize{0.2}
\def\fieldlabelsep{8pt}
\def\stubfrac{0.4}
\def\dotgap{0.45}

\node[gauge]  (L0) at (0,0) {$1$};
\node[flavor] (L1) at (1.25*\xunit,0.95) {$k$};
\node[flavor] (L2) at (1.25*\xunit,-0.95) {$k$};

\draw[midarrow] (L0.35)  -- (L1.215);
\draw[midarrow] (L2.145) -- (L0.325);

\node at (0.62*\xunit,-1.85) {$\mathcal{W}=0$};

\draw[<->, very thick] (2.35*\xunit,0) -- (3.45*\xunit,0);

\pgfmathsetmacro{\xstart}{4.40*\xunit}
\pgfmathsetmacro{\dx}{1.50*\xunit}
\pgfmathsetmacro{\stub}{\stubfrac*\dx}
\pgfmathsetmacro{\nodeoffset}{0.45}

\node[flavor] (R0) at (\xstart,0)             {$1$};
\node[gauge]  (R1) at ({\xstart+\dx},0)       {$1$};
\node[gauge]  (R2) at ({\xstart+2*\dx},0)     {$1$};

\coordinate (A1) at ($(R2.east)+(0.05,0.12)$);
\coordinate (A2) at ($(A1)+(\stub,0)$);

\node (Dots) at ($(A2)+(\dotgap,-0.12)$) {$\cdots$};

\coordinate (B1) at ($(Dots)+(\dotgap,0.12)$);
\coordinate (B2) at ($(B1)+(\stub,0)$);

\node[gauge]  (Rk)   at ($(B2)+(\nodeoffset,-0.12)$) {$1$};
\node[flavor] (Rkp1) at ($(Rk)+(\dx,0)$)              {$1$};

\draw[decorate,decoration={brace,amplitude=5pt}]
  ($(R1.north)+(0,0.65)$) -- ($(Rk.north)+(0,0.65)$)
  node[midway,above=6pt] {$k-1$};

\draw[midarrow]
  ($(R0.east)+(0.05,0.12)$) --
  ($(R1.west)+(-0.05,0.12)$)
  node[phiLabel]
  {$\Phi_{(0,1)}$};

\draw[midarrow]
  ($(R1.west)+(-0.05,-0.12)$) --
  ($(R0.east)+(0.05,-0.12)$)
  node[tphiLabel]
  {$\tilde{\Phi}_{(0,1)}$};

\coordinate (M1) at ($(R0)!0.5!(R1)$);
\draw ($(M1)+(-\crosssize,\crosssize)$) -- ($(M1)+(\crosssize,-\crosssize)$);
\draw ($(M1)+(-\crosssize,-\crosssize)$) -- ($(M1)+(\crosssize,\crosssize)$);

\draw[midarrow]
  ($(R1.east)+(0.05,0.12)$) --
  ($(R2.west)+(-0.05,0.12)$)
  node[phiLabel]
  {$\Phi_{(1,2)}$};

\draw[midarrow]
  ($(R2.west)+(-0.05,-0.12)$) --
  ($(R1.east)+(0.05,-0.12)$)
  node[tphiLabel]
  {$\tilde{\Phi}_{(1,2)}$};

\coordinate (M2) at ($(R1)!0.5!(R2)$);
\draw ($(M2)+(-\crosssize,\crosssize)$) -- ($(M2)+(\crosssize,-\crosssize)$);
\draw ($(M2)+(-\crosssize,-\crosssize)$) -- ($(M2)+(\crosssize,\crosssize)$);

\draw[midarrow] (A1) -- (A2);
\draw[midarrow] (B1) -- (B2);

\coordinate (C1) at ($(Rk.west)+(-0.05,-0.12)$);
\coordinate (C2) at ($(C1)+(-\stub,0)$);
\draw[midarrow] (C1) -- (C2);

\coordinate (D2) at ($(R2.east)+(0.05,-0.12)$);
\coordinate (D1) at ($(D2)+(\stub,0)$);
\draw[midarrow] (D1) -- (D2);

\draw[midarrow]
  ($(Rk.east)+(0.05,0.12)$) --
  ($(Rkp1.west)+(-0.05,0.12)$)
  node[phiLabel]
  {$\Phi_{(k-1,k)}$};

\draw[midarrow]
  ($(Rkp1.west)+(-0.05,-0.12)$) --
  ($(Rk.east)+(0.05,-0.12)$)
  node[tphiLabel]
  {$\tilde{\Phi}_{(k-1,k)}$};

\coordinate (Mk) at ($(Rk)!0.5!(Rkp1)$);
\draw ($(Mk)+(-\crosssize,\crosssize)$) -- ($(Mk)+(\crosssize,-\crosssize)$);
\draw ($(Mk)+(-\crosssize,-\crosssize)$) -- ($(Mk)+(\crosssize,\crosssize)$);

\node[align=center] at ($(R0)!0.5!(Rkp1)+(0,-1.5)$)
{$\mathcal{W}=\sum_{i=1}^{k} F_i\,\Phi_{(i-1,i)}\,\tilde{\Phi}_{(i-1,i)}$};

\end{tikzpicture}
\caption{The 3d abelian mirror pair between SQED with $k$ flavors and the affine $A_{k-1}$ quiver. Circles denote $U(1)$ gauge nodes, while squares denote flavor symmetries with the number inside each node indicating the rank. Arrows denote chiral multiplets in the corresponding bifundamental representations, with the two oppositely oriented arrows between adjacent nodes representing the pair $\Phi_{(i-1,i)}$, $\tilde\Phi_{(i-1,i)}$ with $i=1,\cdots,k-1$. The crosses on the right quiver denote the gauge singlet chirals $F_i$.}
\label{fig:3dmirrquiver}
\end{figure}

Our main example will be the 3d abelian mirror duality for supersymmetric quantum electrodynamics (SQED) with $k$ flavors. In its $\mathcal{N}=4$ formulation, the flavors consist of hypermultiplets of unit charge and the mirror dual theory is given by a quiver theory that takes the form of the affine $A_{k-1}$ Dynkin diagram \cite{Intriligator:1996ex,Hanany:1996ie,deBoer:1997kr}. However, one can also consider the $\mathcal{N}=2$ version of this mirror duality \cite{Aharony:1997bx}, which has the advantage of making a larger flavor symmetry group manifest. We schematically depict this duality in a quiver notation in Figure \ref{fig:3dmirrquiver}. 

The duality for generic $k$ can be obtained by iterating the one for $k=1$ using a piecewise procedure due to Kapustin and Strassler \cite{Kapustin:1999ha}. This is one of the first instances of deriving a duality from another one, working in a fixed number of dimensions. The basic idea behind such a derivation is to view the $k=1$ duality as a functional Fourier transform at the level of the path integral. One of our main results will be a two-dimensional version of Kapustin--Strassler, which will allow us to prove the identities that we will get from the small circle limit of the 3d abelian mirror index identities.

In Section \ref{sec:3d} we review the 3d abelian mirror duality, the logic of Kapustin--Strassler, and its implementation at the level of the supersymmetric index discussed in \cite{Kapustin:2011jm}. The latter allows us to derive the index identity for the generic $k$ duality by assuming a small generalization of the one for $k=1$ with a background flux $n\in\mathbb{Z}$ for the topological symmetry of SQED turned on.

In Section \ref{sec:SQED/XYZ} we begin our systematic study of the limits of the index identities for such abelian mirror dualities. We start from the $\mathcal{N}=2$ variant of the $k=1$ duality, which is often referred to as the SQED/XYZ duality, since the dual theory is just a Wess--Zumino (WZ) model of three chiral fields with no gauge group. We will denote by Limit 1 and Limit 2 the two possible scalings of parameters that we introduced before from the perspective of SQED (remember that the role of these parameters is swapped in the dual theory). 

From Limit 1, we get the identity
\begin{align}\label{eq:id1}
        \frac{(-1)^n}{\pi}  \int_{\mathbb{C}}  \frac{\mathrm{d}^2Z}{Z^{\frac{3}{4}-\frac{n}{2}+\phi+\tau}\bar{Z}^{\frac{3}{4}+\frac{n}{2}+\phi+\tau}\left|1-Z\right|^{1-4\phi}}=\frac{\Gamma(\frac{1}{2}+2\phi)\Gamma(\frac{1}{4}\pm \frac{n}{2}-\phi\pm\tau)}{\Gamma(\frac{1}{2}-2\phi)\Gamma(\frac{3}{4}\pm \frac{n}{2}+\phi\mp\tau)}\,.
\end{align}
On the left we have the result of the $\beta\to0$ limit of the index of SQED. It can be interpreted either as the $S^2$ partition function of a LG model, or as a CFT correlator. On the right instead we have the limit of the index of the XYZ theory, which gives the partition function of a 2d theory with the same field content of three chirals. In the first interpretation, we recover a duality that can be derived from the simplest example of 2d Hori--Vafa duality for SQED with one flavor by gauging the magnetic $(-1)$-form symmetry. This gauging can be viewed as a 2d version of the functional Fourier transform of Kapustin--Strassler. In the second case, we obtain a standard evaluation formula for the Coulomb gas integral of the three-point function of Liouville CFT with one screening charge, which is essentially a specific residue of the more general DOZZ formula \cite{Dorn:1994xn,Zamolodchikov:1995aa}. This is also a particular case of a similar result discussed in \cite{Pasquetti:2019uop}. The origin of the various parameters is explained in the main text, but we point out that the extra parameter $n$ is an integer that descends from the background flux for the topological symmetry of SQED. Its refinement is crucial for the 2d version of Kapustin--Strassler. The identity for $n=0$ is actually a very standard integral representation of the complex beta-function, which also appears in the study of closed-string amplitudes as the Virasoro--Shapiro amplitude for the scattering of four tachyons in flat space \cite{Virasoro:1969me,Shapiro:1970gy}.

From Limit 2 instead, we get the identity
\begin{equation}\label{eq:id2}
    \sum_{m\in \mathbb{Z}}\int_{-\infty}^{\infty}\mathrm{d}\sigma (-\xi)^{m} \frac{\Gamma (\frac{1}{4}+\phi\pm (\frac{m}{2}+2\pi i \sigma))}{\Gamma (\frac{3}{4}-\phi\pm (\frac{m}{2}-2\pi i \sigma))}=(1-\xi)^{-\frac{1}{2}-2\phi}(1-\xi^{-1})^{-\frac{1}{2}-2\phi}\frac{\Gamma(\frac{1}{2}+2\phi)}{\Gamma (\frac{1}{2}-2\phi)}\,.
\end{equation}
The index of SQED reduces to the partition function of a similar SQED theory with one flavor in 2d, while the index of the XYZ model reduces to the partition function of the mirror dual LG model. This is a degenerate case from the CFT perspective, as it will be clearer when discussing the higher $k$ generalization. We also explain how the Limit 2 identity can be derived from the Limit 1 identity, showing that they are not independent from each other. 

In Section \ref{sec:limitk} we generalize this analysis to higher number $k$ of flavors on the SQED side. Again we separate the discussion for Limit 1 and Limit 2, and provide the two possible physical interpretations in each case. From Limit 1, we get the identity
\begin{align}\label{eq:id3}
    \begin{split}
         &\frac{1}{\pi}\int_{\mathbb{C}}\mathrm{d}^2Z \frac{1}{|Z|^{2+2\tau}}\prod_{i=1}^k\left[\frac{1}{|v_iZ|^{-\frac{1}{2}+2\phi_i}|1-Zv_i|^{1-4\phi_i}}\right]=\prod_{i=1}^k\left[\frac{\Gamma(\frac{1}{2}+2\phi_i)}{\Gamma(\frac{1}{2}-2\phi_i)}\right]\\
         &\times\sum_{\vec{m}\in\mathbb{Z}^{k-1}} \prod_{a=1}^{k-1}\left[\int_{-\infty}^\infty\mathrm{d}\sigma_a\left(\frac{v_a}{v_{a+1}}\right)^{-m_a}\right]\prod_{i=1}^{k}\left[\frac{\Gamma(\frac{1}{4}-\phi_i\pm(\tau\delta_{i,k}+\frac{m_i-m_{i-1}}{2}+2\pi i (\sigma_{i-1}-\sigma_i)) }{\Gamma(\frac{3}{4}+\phi_i\pm(\tau\delta_{i,k}+\frac{m_{i-1}-m_{i}}{2}+2\pi i (\sigma_{i-1}-\sigma_i)) }\right]\,.
    \end{split}
\end{align}
This is an equality between the partition function of an abelian GLSM that is identical to the $A_{k-1}$ quiver we started with in 3d on the right, and the partition function of its mirror dual LG model or equivalently the Coulomb gas integral with one screening charge for the $(k+2)$-point function of Liouville CFT on the left. 

From Limit 2 instead, we get the identity
\begin{align}\label{eq:id4}
    &\sum_{m\in \mathbb{Z}}\int_{-\infty}^{\infty}\mathrm{d}\sigma (-1)^{km}\xi^m\prod_{i=1}^k\frac{\Gamma \left(\frac{1}{4}+\phi_i\pm \left(\frac{m}{2}+2\pi i (\sigma+\nu_i)\right)\right)}{\Gamma \left(\frac{3}{4}-\phi_i\pm \left(\frac{m}{2}-2\pi i (\sigma+\nu_i)\right)\right)}\nn\\ 
    &\qquad=\prod_{i=1}^k\left[\frac{\Gamma(\frac{1}{2}+2\phi_i)}{\Gamma(\frac{1}{2}-2\phi_i)}\right] \prod_{a=1}^{k-1}\left[\int_{\mathbb{C}} \frac{\mathrm{d}^2Z_a}{\pi |Z_a|^2}\frac{|Z_a|^{-4\pi i (\nu_a-\nu_{a+1})+2(\phi_a-\phi_{a+1})}}{|1-\frac{Z_a}{Z_{a-1}}|^{1+4\phi_a}}\right]\frac{1}{|1-\frac{1}{Z_{k-1}\xi}|^{1+4\phi_k}}\,.
\end{align}
This is an equality between the partition function of SQED with $k$ flavors, and the partition function of its mirror dual LG model or equivalently the Coulomb gas integral with one screening charge for a four-point function of $A_{k-1}$ Toda CFT with two arbitrary, one semi-degenerate and one degenerate vertex operators. From the latter perspective, it is clear that the CFT interpretation is degenerate for $k=1$, as mentioned before. The equality of the partition functions of the mirror dual theories has been previously proved in \cite{Gomis:2012wy}. Here we show how to obtain the same identity from the 3d perspective, and also how to manipulate the LG partition function into the Toda Coulomb gas integral. Moreover, the fact that the partition function of 2d $\mathcal{N}=(2,2)$ SQED with $k$ flavors equates the four-point function of Toda was already observed in \cite{Doroud:2012xw}, but here we make direct contact with the Coulomb gas integral form of the Toda correlator, which is relevant for the 2d/2d correspondence.

All of the above identities have analogs where one considers the hemisphere rather than sphere partition functions of the 2d $\mathcal{N}=(2,2)$ theories, and the conformal blocks rather than full correlators in the CFT. These versions of the identities typically correspond to various integral representations of hypergeometric functions. For example, the hemisphere/conformal block analog of the identity \eqref{eq:id1} is just the standard integral representation of the beta-function, while for \eqref{eq:id2} we have the Mellin--Barnes integral representation of the ${}_1F_0$ hypergeometric function. The two are related by a Mellin transform. Instead, the hemisphere/conformal block analog of \eqref{eq:id4} corresponds to two different integral representations of the ${}_kF_{k-1}$ hypergeometric function, the Mellin--Barnes representation on the left and the Euler one on the right. The variant of \eqref{eq:id3} does not correspond to any hypergeometric function, but is still related to Selberg/Aomoto-type integrals.

In Section \ref{sec:2dKS} we propose a two-dimensional version of the Kapustin--Strassler piecewise procedure. This provides an independent proof of both the Limit 1 and Limit 2 identities for generic $k$ \eqref{eq:id3} and \eqref{eq:id4} by just iterating the Limit 1 identity for $k=1$ \eqref{eq:id1}. Hence, the latter is the only actually fundamental identity, since all the others can be derived from it.

In Section \ref{sec:2d2dAD} we provide an example where the identity \eqref{eq:id4} can be applied in the context of the 2d/2d correspondence of \cite{Rastelli:2025nyn}. We consider as 4d $\mathcal{N}=2$ SCFT one of the Argyres--Douglas theories $T=(A_{k-1},A_{N-1})$ with $\mathrm{gcd}(k,N)=1$ \cite{Argyres:1995jj,Argyres:1995xn,Xie:2012hs}. These theories possess the property of having a trivial Higgs branch, which reflects the fact that the associated VOA is the $(k,k+N)$ $\mathcal{W}_k$ algebra \cite{Cordova:2015nma,Xie:2016evu}. Hence, following \cite{Rastelli:2025nyn}, we expect that the full 2d CFT that we obtain from $S^2$ compactification of $(A_{k-1},A_{N-1})$ should be the non-unitary $(k,k+N)$ $W_k$ minimal model. The four-point functions of the latter can be obtained as a rational specialization of those of Toda. Restricting to those that admit the Coulomb gas integral representation in \eqref{eq:id4}, we can then use this identity to find a GLSM description of the 2d (2,2) theory whose $S^2$ partition function reproduces the minimal model correlator. We interpret the latter as descending from the $SU(2)_R$-twisted compactification of $(A_{k-1},A_{N-1})$ on a four-punctured sphere, where the type of punctures is related to the choice of primaries in the CFT correlator. This generalizes the analysis of \cite{Rastelli:2025nyn} for the Virasoro minimal models $k=2$.

A comment about the domain of the parameters in the integral identities that we have just reviewed is in order. The identities will be initially derived under the assumption that
\begin{equation}\label{eq:reality}
    \tau,\,\phi_i \in i\mathbb{R}\,,\qquad \nu_i\in\mathbb{R}\,,\qquad |\xi|=|v_i|=1\,.
\end{equation}
This has a natural interpretation from the perspective of the 2d (2,2) GLSMs. Indeed, the parameters $\tau$, $\phi_i$, $\nu_i$ correspond (up to an $i$ factor for the first two due to our conventions) to real masses that we can turn on in the partition function for each $U(1)$ factor in the Cartan of the flavor symmetry. Instead, $\xi$ and $v_i$ are identified up to a complex exponential map to the FI parameters which are also real.

However, all of the integral identities can generically be analytically continued so that these parameters can be more generic complex numbers. For the mass parameters, this is thanks to the fact that the R-symmetry can be mixed with each $U(1)$ flavor symmetry. As explained in \cite{Benini:2012ui,Doroud:2012xw}, the masses are then promoted to complex numbers, where the imaginary part is given by the mixing coefficients. Schematically
\begin{equation}\label{eq:mixing}
    M=\tilde{m}+i\frac{r}{4\pi}\,,
\end{equation}
where $\tilde{m}$ is the real mass and $r$ is the mixing coefficient of the associated flavor symmetry with the R-symmetry. Not any value of the mixing coefficients is however allowed in order for the integrals to converge. The minimal criterion, which is what we will always assume in the present paper, is that all the chiral multiplets of the GLSM have an R-charge between 0 and 1 (in units where the superpotential has R-charge 2). For example for \eqref{eq:id3} this means
\begin{equation}
    \text{Domain of \eqref{eq:id3}}:\quad 0<2\mathrm{Re}(\phi_i)<1\,,\qquad 0<\frac{1}{2}-\mathrm{Re}(\phi_i)\pm\delta_{i,k}\mathrm{Re}(\tau)<1\,,
\end{equation}
while for \eqref{eq:id4} we have
\begin{equation}
    \text{Domain of \eqref{eq:id4}}:\quad 0<2\mathrm{Re}(\phi_i)<1\,,\qquad 0<\mathrm{Re}(\phi_i)\pm2\pi\mathrm{Im}(\nu_i)<1\,.
\end{equation}
In many cases, one can perform a further analytic continuation to a larger domain. For example, for $k=1$ we can see from the r.h.s.~of \eqref{eq:id1} and \eqref{eq:id2} that we can consider more general values of $\mathrm{Re}(\phi)$ and $\mathrm{Re}(\tau)$ as long as the arguments of the gamma functions are not negative integers. We will refrain from a more in-depth and systematic analysis of how far the analytic continuation can be pushed here since it is beyond the scope of this paper. 

For the FI parameters instead, these can be promoted to be complex by turning on a theta angle
\begin{equation}
    t=\frac{\theta}{2\pi}+i\eta\,,
\end{equation}
where $\eta$ is the FI and $\theta$ the theta angle.

This discussion has a complete parallel in three dimensions. In the supersymmetric index we have parameters called fugacities that are usually taken to be pure phases
\begin{equation}
    f=x^{-2i\tilde{m}}=\mathrm{e}^{i\beta\tilde{m}}\,.
\end{equation}
where $x=\mathrm{e}^{-\frac{\beta}{2}}$. From this perspective, $\tilde{m}$ is the holonomy of the background gauge field for the associated global symmetry around the $S^1$ and is thus real. This is the assumption that gives the reality conditions \eqref{eq:reality} of the 2d parameters, up to the identification of 3d and 2d parameters that we will explain in more detail in the main text and that is related to the Limits 1 and 2.

Also in this case, we can promote the fugacities to arbitrary complex numbers by mixing with the R-symmetry. In particular, \eqref{eq:mixing} is achieved in 3d by rescaling the fugacity with a suitable power of the R-symmetry fugacity $x$
\begin{equation}
    f\to x^{\frac{r}{2\pi}}f=\mathrm{e}^{i\beta M}\,.
\end{equation}
Note that in 3d the FI can also be viewed as a background gauge field for a global symmetry, the topological symmetry. Hence, the associated fugacity is on the same footing as the flavor fugacities and can acquire a non-trivial modulus with the same phenomenon of mixing the R-symmetry with the topological symmetry.

\subsection*{Future directions}

The main natural follow up to our work is to consider non-abelian mirror dualities in 3d and the small circle limit of the associated index identities. Many examples of such dualities are known, and the analysis that we have performed here for the abelian dualities should apply to those cases as well, without any substantial modification. Considering non-abelian mirror pairs would allow us to:
\begin{itemize}
    \item obtain sphere partition function identities for non-abelian 2d $\mathcal{N}=(2,2)$ mirror dualities, which have been less explored than the abelian ones (see \cite{Hori:2000kt,Gu:2018fpm,Gu:2019zkw,Gomis:2012wy} for some works);
    \item obtain identities between partition functions of non-abelian GLSMs and Coulomb gas integrals of Toda or Liouville CFT with multiple screening charges (see \cite{Nedelin:2017nsb,Rastelli:2025nyn} for some examples).
\end{itemize}

The main limitation to these explorations is that, unlike the abelian case, there is no simple way to deform the $\mathcal{N}=4$ non-abelian dualities to $\mathcal{N}=2$. Mirror duals of 3d $\mathcal{N}=2$ theories have been discussed in \cite{Giacomelli:2017vgk,Benvenuti:2023qtv,Benvenuti:2024seb,Benvenuti:2025huk,Benvenuti:2026usm,Benvenuti:2026xcv}, however their structure is more complicated than their $\mathcal{N}=4$ counterparts. In particular, the mirror dual of the 3d $\mathcal{N}=2$ non-chiral SQCD of \cite{Benvenuti:2023qtv} is a quasi-Lagrangian theory, since it is obtained by gauging some non-Lagrangian building blocks \cite{Gaiotto:2008ak,Pasquetti:2019tix,Pasquetti:2019hxf,Benvenuti:2024mpn}. The simplest approach would then be to consider the reduction of the $\mathcal{N}=4$ dualities. However, these typically display a smaller global symmetry. As a consequence, the identities that one would obtain in the $\beta\to0$ limit would be refined by fewer parameters than what one might hope. In particular, from the CFT perspective, some combinations of the momenta of the vertex operators would be fixed and so we would not be able to recover the most general expression of the correlator for completely arbitrary values of the momenta. Nevertheless, this might not be a problem from the perspective of the 2d/2d correspondence, since the momenta are often already fixed to specific values in that context.

It would also be interesting to explore whether a 2d version of the mirror dualization algorithm \cite{Hwang:2021ulb,Bottini:2021vms,Comi:2022aqo} exists. This algorithm is conceptually a generalization of the Kapustin--Strassler procedure to non-abelian mirror dualities. It consists of some basic duality moves that are applied sequentially to a given quiver theory to derive its mirror dual, in a way that is very similar to the manipulations on the Type IIB brane setups for such 3d $\mathcal{N}=4$ theories \cite{Hanany:1996ie}. The main difficulty here is that these duality moves are more complicated than those of the Kapustin--Strassler piecewise procedure, since they involve the non-Lagrangian building blocks.

Another direction is to investigate further the connection with the 2d/2d correspondence. In Section \ref{sec:2d2dAD} we have used some of our results to determine GLSMs whose partition functions coincide with certain $W_k$ minimal models four-point functions, and which can thus be associated to $SU(2)_R$-twisted reduction on a four-punctured sphere of the 4d Argyres--Douglas theories. However, as explained in \cite{Rastelli:2025nyn}, there are more checks of this claim that one can perform. One consists of comparing the central charges of the GLSMs with those predicted from 4d. The main difficulty here is figuring out the contribution by anomaly inflow from the punctures, since it requires studying boundary conditions for the 4d theories on a finite radius circle. Properties of the 4d theories on a finite radius circle are well understood for the $(A_1,A_{2M})$ case, see e.g.~\cite{Fredrickson:2017yka}, but not for the generic $(A_{k-1},A_{N-1})$ case. 

Another consistency check is based on a second correspondence proposed in \cite{Rastelli:2025nyn}, which relates the elliptic genus of the same class $\FF$ theories obtained from $SU(2)_R$-twisted compactification on $\Sigma$ of the 4d $\mathcal{N}=2$ SCFT, and correlation functions of a TQFT on $\Sigma_{g,n}$. This is analogous to the superconformal index of class $\SS$ theory being computed by a 2d TQFT \cite{Gadde:2009kb,Gadde:2011ik,Gadde:2011uv}. In \cite{Rastelli:2025nyn} the structure constants and the propagators of this 2d TQFT have been computed for the cases of $(A_1,A_{2M})$. It would be worthwhile to perform a similar analysis for $(A_{k-1},A_{N-1})$, however this would probably require studying more four-point functions of the $(k,k+N)$ $W_k$ minimal models beyond those considered here.

Finally, higher-point functions on a sphere of the minimal models also admit a Coulomb gas integral representation. Finding GLSMs whose partition functions reproduce these Coulomb gas integrals would allow us to access the compactifications of the Argyres--Douglas theories on spheres with more than four punctures. These Coulomb gas integrals typically involve multiple screening charges (see \cite{Rastelli:2025nyn} for the $(A_1,A_2)$ case). It is thus crucial for this to investigate the identities that come from the small circle limit of the 3d index identities associated with non-abelian mirror dualities.


\section{Review of 3d abelian mirror dualities}
\label{sec:3d}

We consider the 3d $\mathcal{N}=2$ duality between SQED with $k$ flavors and the $A_{k-1}$ abelian quiver \cite{Intriligator:1996ex,Hanany:1996ie,Aharony:1997bx,deBoer:1997kr}, which is also summarized with quivers in Figure \ref{fig:3dmirrquiver}:
\begin{itemize}
    \item \textbf{Theory A:} $U(1)$ gauge theory with $k$ pairs of chirals $\Phi_i$, $\tilde{\Phi}_i$ of charge $\pm1$ and no superpotential.
    \item \textbf{Theory B:} $\prod_{a=1}^{k-1}U(1)_a$ gauge theory with $k$ pairs of chirals $\Phi_{(i-1,i)}$, $\tilde{\Phi}_{(i-1,i)}$ charged under it as in Table \ref{tab3ddualitycharges} and $k$ gauge singlet chiral fields $F_i$ interacting with the superpotential
    \begin{equation}
    \mathcal{W}=\sum_{i=1}^kF_i\Phi_{(i-1,i)}\tilde{\Phi}_{(i-1,i)}\,.
    \end{equation}
\end{itemize}

\begin{table}[t]
\centering
\begin{tabular}{c|cccc}
    & $U(1)_a$ & $U(1)_{s_i}$ & $U(1)_\xi$ & $U(1)_R$ \\\hline
    $\Phi_{(0,1)}$ & $-\delta_{a,1}$ & $-\delta_{i,1}$ & 0 & $\frac{1}{2}$ \\
    $\tilde{\Phi}_{(0,1)}$ & $\delta_{a,1}$ & $-\delta_{i,1}$ & $0$ & $\frac{1}{2}$ \\
    $\Phi_{(i-1,i)}$, $i=2,\cdots,k-1$ & $-\delta_{a,i}+\delta_{a+1,i}$ & $-1$ & $0$ & $\frac{1}{2}$ \\
    $\tilde{\Phi}_{(i-1,i)}$, $i=2,\cdots,k-1$ & $\delta_{a,i}-\delta_{a+1,i}$ & $-1$ & $0$ & $\frac{1}{2}$ \\
    $\Phi_{(k-1,k)}$ & $\delta_{a,k-1}$ & $-\delta_{i,k}$ & 1 & $\frac{1}{2}$ \\
    $\tilde{\Phi}_{(k-1,k)}$ & $-\delta_{a,k-1}$ & $-\delta_{i,k}$ & $-1$ & $\frac{1}{2}$ \\
    $F_i$, $i=1,\cdots,k$ & 0 & 2 & 0 & $1$
\end{tabular}
\caption{Charges under the gauge and global symmetries of the chiral fields of Theory B. We denote by $U(1)_R$ a convenient choice of R-symmetry, which is not necessarily the superconformal one.}
\label{tab3ddualitycharges}
\end{table}

The global symmetry of the low energy theory is fully manifest in Theory A
\begin{equation}
    SU(k)_v\times U(k)_s\times U(1)_{\xi}\,,
\end{equation}
where we are labelling each symmetry by the fugacities we will use in the index. From the perspective of Theory A, $SU(k)_v$ is the vector symmetry (whose $U(1)$ part is gauged), $U(k)_s$ is the axial symmetry, and $U(1)_{\xi}$ is the topological symmetry.\footnote{More precisely, the fields $\Phi_i$ form the fundamental representation of a $U(k)_f$ global symmetry and the fields $\tilde{\Phi}_i$ the fundamental of $U(k)_{\tilde{f}}$. We then define fugacities for the vector and axial symmetries as
\begin{equation}
    v_i=(f_i\tilde{f}_i^{-1})^{\frac{1}{2}}\,,\qquad s_i=(f_i\tilde{f}_i)^{\frac{1}{2}}
\end{equation}
and $u=\prod_i v_i$ is the gauge fugacity.} From the perspective of Theory B instead, $SU(k)_v$ is enhanced from the $k-1$ abelian topological symmetries of each gauge node, $U(k)_s$ is enhanced from the $\prod_{i=1}^{k}U(1)_{s_i}$ axial symmetries corresponding to each $\Phi_{(i,i+1)}$, $\tilde{\Phi}_{(i,i+1)}$ pair, and $U(1)_{\xi}$ is the vector symmetry (see Table \ref{tab3ddualitycharges}).

The supersymmetric index identity associated to this duality is \cite{Imamura:2011su,Kapustin:2011jm,Krattenthaler:2011da} (we spell out our conventions in Appendix \ref{app:SUSYpf})
\begin{align}
    \label{eq:U1Index}
        \mathcal{I}_{\text{A}}&=\sum_{m\in\mathbb{Z}}\xi^m\oint\frac{\mathrm{d}u}{2\pi i u}\left(-u\right)^{km}  \prod_{i=1}^k\frac{((uv_i)^{\mp1}s_i^{-1}x^{\frac{3}{2}\mp m};x^2)_{\infty}}{((uv_i)^{\pm1}s_ix^{\frac{1}{2}\mp m};x^2)_{\infty}}\nn\\
        &=\prod_{i=1}^k\left[\frac{\left(s^{-2}_{i}x;x^2\right)}{\left(s^{2}_{i}x;x^2\right)} \right]\sum_{\vec{m}\in\mathbb{Z}^{k-1}}\xi^{m_{k-1}}\oint \prod_{a=1}^{k-1} \left[\frac{\mathrm{d}u_a}{2\pi i u_a}\left(\frac{v_a}{v_{a+1}}\right)^{-m_a}u_a^{-m_{a-1}+2m_a-m_{a+1}}\right]\nn\\
        &\times\prod_{i=1}^k\left[\frac{\left(\left(\frac{u_i}{u_{i-1}})\right)^{\pm1}s_i\xi^{\mp\delta_{i,k}}x^{\frac{3}{2}\pm(m_i-m_{i-1})};x^2\right)_\infty}{\left(\left(\frac{u_i}{u_{i-1}})\right)^{\mp1}s_i^{-1}\xi^{\pm\delta_{i,k}}x^{\frac{1}{2}\pm(m_i-m_{i-1})};x^2\right)_\infty} \right]=\mathcal{I}_{\text{B}}\,.
\end{align}
In our notation, the fugacities for the $SU(k)_v$ vector symmetry satisfy the constraint $\prod_{i=1}^kv_i=1$, and we also introduced dummy variables $u_0=u_k=1$ and $m_0=m_k=0$ to write the expressions in a more compact form. The fugacity $x$ can be understood as counting the combination $R+2J_3$ of the R-charges and spins of the states of the theory, but if we think of the index as a partition function on $S^2\times S^1$ it is also identified as $x=\mathrm{e}^{-\frac{\beta}{2}}$, where $\beta$ is the ratio of the radii of $S^1$ and $S^2$. The limit that we are interested in is $\beta\to0$, in which the circle shrinks to zero.

The simplest duality among this class corresponds to $k=1$. In this case, we have that SQED with one flavor is dual to a WZ model of three chiral fields with a cubic superpotential (also known as the XYZ model)
\begin{itemize}
    \item \textbf{Theory A:} $U(1)$ gauge theory with one pair of chirals $\Phi$, $\tilde{\Phi}$ of charge $\pm1$ and no superpotential.
    \item \textbf{Theory B:} WZ model of three chirals $X$, $Y$, $Z$ with cubic superpotential
    \begin{equation}
    \mathcal{W}=XYZ\,.
    \end{equation}
\end{itemize}
It will turn out to be useful for us to consider a refinement of the associated index identity \eqref{eq:U1Index} for $k=1$, which corresponds to turning on a background magnetic flux $n$ for the $U(1)_\xi$ symmetry \cite{Kapustin:2011jm}
\begin{align}
    \label{eq:U1Indexk1}
        \sum_{m\in\mathbb{Z}}\oint\frac{\mathrm{d}u}{2\pi i u}\left(-u\xi\right)^{m+n}  \frac{(u^{\mp1}s^{-1}x^{\frac{3}{2}\mp m};x^2)_{\infty}}{(u^{\pm1}sx^{\frac{1}{2}\mp m};x^2)_{\infty}}=\frac{\left(s^{-2}x;x^2\right)\left(s\xi^{\mp1}x^{\frac{3}{2}\pm n};x^2\right)}{\left(s^{2}x;x^2\right)\left(s^{-1}\xi^{\pm1}x^{\frac{1}{2}\pm n};x^2\right)}\,.
\end{align}

Kapustin and Strassler \cite{Kapustin:1999ha} pointed out that this SQED/XYZ duality can be considered as a functional Fourier transform at the level of the path integral. To understand this, it is convenient to look at the $\mathcal{N}=4$ version of the duality, where one of the fields of the WZ model is moved to the SQED side and interpreted as the $\mathcal{N}=2$ chiral inside the $\mathcal{N}=4$ vector multiplet. The resulting duality is
\begin{equation}\label{eq:N4duality}
    \text{3d }\mathcal{N}=4\text{ SQED with one hyper}\,\,\longleftrightarrow\,\,\text{one free hyper}\,.
\end{equation}
The l.h.s.~consists of the $U(1)$ gauging of a hyper with a BF coupling to the FI parameter
\begin{equation}\label{eq:BF}
\mathcal{L}_{\text{BF}}=\int \mathrm{d}^2\theta\mathrm{d}^2\tilde{\theta}\,\left[\Sigma\,V_{\text{FI}}+h.c.\right]\,,
\end{equation}
where $V_{\text{FI}}$ is the background field for the topological symmetry whose bottom component gives the FI parameter, while $\Sigma$ is the field strength multiplet. The idea is then to view this gauging as a functional Fourier transform and the duality as the property of the free hyper theory of being invariant under it.

This statement is more neatly understood at the level of the $S^3$ partition function \cite{Kapustin:2009kz,Kapustin:2010xq}
\begin{equation}\label{eq:S3pf}
    \underbrace{\int_{-\infty}^{\infty}\mathrm{d}\sigma}_{U(1)\text{ gauging}}\underbrace{\mathrm{e}^{2\pi i\eta\sigma}}_{\text{FI}}\underbrace{\frac{1}{\cosh{(\pi\sigma)}}}_{\text{hyper}}=\frac{1}{\cosh{(\pi\eta)}}\,,
\end{equation}
where we have highlighted the origin of each contribution. We then see that the partition function of a hypermultiplet is invariant under the ordinary Fourier transform. The identity \eqref{eq:U1Indexk1} is a different manifestation of the same idea at the level of the index, where the $U(1)$ gauging is implemented by an integral over the gauge holonomy $u$ and a sum over the gauge magnetic flux $m$.

The procedure that we have just described can be reversed. As mentioned before, the l.h.s.~of \eqref{eq:N4duality} consists of gauging the $U(1)$ flavor symmetry of a free hyper. In the process, we gain a new $U(1)$ topological symmetry, whose background field encodes the FI parameter. This is achieved by introducing the BF coupling \eqref{eq:BF}. We can at this point undo this process by further gauging the $U(1)$ topological symmetry. At the level of the $S^3$ partition function, this amounts to integrating over $\eta$
\begin{equation}
    \int_{-\infty}^{\infty}\mathrm{d}\eta\,\mathrm{e}^{2\pi i\omega\eta}\int_{-\infty}^{\infty}\mathrm{d}\sigma\frac{\mathrm{e}^{2\pi i\eta\sigma}}{\cosh{(\pi\sigma)}}=\int_{-\infty}^{\infty}\mathrm{d}\sigma\frac{\delta(\sigma-\omega)}{\cosh{(\pi\sigma)}}\,.
\end{equation}
We immediately see that the BF coupling induces a delta-function that can be used to remove the original $U(1)$ gauging and get back the theory of a free hyper
\begin{equation}
        \int_{-\infty}^{\infty}\mathrm{d}\sigma\frac{\delta(\sigma-\omega)}{\cosh{(\pi\sigma)}}=\frac{1}{\cosh{(\pi\omega)}}\,.
\end{equation}
In other words, the functional Fourier transform is an invertible process and the inverse operation consists of gauging the topological symmetry. 

Performing the same operation on both sides of the duality we get back a similar identity to \eqref{eq:S3pf}
\begin{equation}
    \frac{1}{\cosh{(\pi\sigma)}}=\int_{-\infty}^{\infty}\mathrm{d}\eta\frac{\mathrm{e}^{2\pi i\omega\eta}}{\cosh{(\pi\eta)}}\,.
\end{equation}
The same happens at the level of the index, as it can be seen by integrating over $\xi$ and summing over $n$ on both sides of \eqref{eq:U1Indexk1}. This shows that the basic abelian mirror duality for SQED with one flavor is invariant under the gauging process.

\subsection*{Kapustin--Strassler derivation at the index level}

Kapustin and Strassler \cite{Kapustin:1999ha} have also shown that this functional Fourier transform perspective on the basic abelian mirror duality for SQED with one flavor can be used to derive the more general duality for any number $k$ of flavors. In other words, they proposed a piecewise derivation, which assumes only the duality in the particular case $k=1$ and iterates it to produce the one for generic $k$. This derivation was used at the level of the $S^3$ \cite{Kapustin:2010xq} and $S^3_b$ \cite{Benvenuti:2016wet} partition function, and at the level of the supersymmetric index \cite{Kapustin:2011jm} to prove the integral identities associated to the generic $k$ duality by simply iterating those for $k=1$. Here we will review the index derivation, since it will be instructive to understand the 2d version of Kapustin--Strassler that we will discuss in Section \ref{sec:2dKS}. In other words, we are going to review how we can derive the identity \eqref{eq:U1Index} for general $k$ by just assuming the generalization \eqref{eq:U1Indexk1} with a background flux for the topological symmetry of the $k=1$ identity.

The derivation consists of three main steps:
\begin{enumerate}
    \item We start from the index of SQED with $k$ flavors that is given on the l.h.s.~of \eqref{eq:U1Index} and trade the contribution of each flavor for an auxiliary pair of integral and sum using  \eqref{eq:U1Indexk1} from right to left
    \begin{align}\label{eq:U1IndexKSstep1}
        \mathcal{I}_{\text{A}}&=\sum_{m\in\mathbb{Z}}\oint\frac{\mathrm{d}u}{2\pi i u}\left(-u\right)^{km} \xi^m\prod_{i=1}^k\sum_{l_i\in\mathbb{Z}}\oint\frac{\mathrm{d}w_i}{2\pi i w_i}(-w_iuv_i)^{l_i-m}\frac{(w_i^{\mp1}s_ix^{\frac{3}{2}\mp l_i};x^2)_{\infty}}{(w_i^{\pm1}s_i^{-1}x^{\frac{1}{2}\mp l_i};x^2)_{\infty}}\,.
    \end{align}
    The variables $u$, $m$ do not appear in any $q$-Pochhammer symbol anymore. Physically, this means that we have reached a dual frame with $k$ copies of SQED with one flavor, where each of these auxiliary $U(1)$ gauge symmetries couples to the original $U(1)$ only via mixed Chern--Simons (CS) interactions.
    \item We can evaluate the integral over $u$ and the sum over $m$, leading to constraints on the remaining variables
    \begin{align}
    \sum_{m\in\mathbb{Z}}\oint\frac{\mathrm{d}u}{2\pi i u}(-u)^{\sum_{i=1}^kl_i} \Big(\xi^{-1}\prod_{i=1}^kw_i\Big)^{-m}=\delta\Big(\xi^{-1}\prod_{i=1}^kw_i\Big)\delta_{\sum_{i=1}^kl_i,0}\,.
    \end{align}
   Plugging this into \eqref{eq:U1IndexKSstep1}, we get
    \begin{align}\label{eq:U1IndexKSstep1}
        \mathcal{I}_{\text{A}}&=\prod_{i=1}^k\sum_{l_i\in\mathbb{Z}}\oint\frac{\mathrm{d}w_i}{2\pi i w_i}\delta\Big(\xi^{-1}\prod_{i=1}^kw_i\Big)\delta_{\sum_{i=1}^kl_i,0}(w_iv_i)^{l_i}\frac{(w_i^{\mp1}s_ix^{\frac{3}{2}\mp l_i};x^2)_{\infty}}{(w_i^{\pm1}s_i^{-1}x^{\frac{1}{2}\mp l_i};x^2)_{\infty}}\,.
    \end{align}
    \item We use the constraints to eliminate an additional pair of integral and sum. We do so by taking
    \begin{align}
        &w_i=u_{i-1}u_i^{-1}\,,\quad i=1,\cdots,k-1\,,\qquad w_k=u_{k-1}u_k^{-1}\xi\,,\nn\\
        &l_i=m_{i-1}-m_i\,,\quad i=1,\cdots,k\,,
    \end{align}
    where $u_0=u_k=1$ and $m_0=m_k=0$. We then recover the index $\mathcal{I}_{\text{B}}$ of the $A_{k-1}$ abelian quiver that is the r.h.s.~of \eqref{eq:U1Index}.
\end{enumerate}


\section{Reduction of the SQED/XYZ duality}
\label{sec:SQED/XYZ}

\subsection{Limit 1 of the index}
\label{SubSection:Limit1k1} 

We are interested in the small circle limit, that is the $\beta\to0$ limit where $x^2=\mathrm{e}^{-\beta}$, of the index identity \eqref{eq:U1Indexk1} associated with the SQED/XYZ duality in the presence of a background magnetic flux $n$ for the $U(1)_\xi$ symmetry. The result of this limit depends on how the fugacities in the index scale with $\beta$. The first possibility that we consider, which we shall refer to as Limit 1, is 
\begin{equation}
\label{eq:Limit1scaling}
    \textbf{Limit 1}:\qquad\xi=x^{2\tau}\,,\qquad s=x^{2\phi}\,,\qquad \tau,\phi\, \,\text{ finite}\,.
\end{equation}
This limit actually corresponds to a particular case of the limit studied in \cite{Pasquetti:2019uop}, which instead considered as starting point a 3d confining duality for a non-abelian $U(N)$ gauge theory. Compared to \cite{Pasquetti:2019uop} we will work with the additional background magnetic flux $n$ turned on.

Following \cite{Pasquetti:2019uop,ArabiArdehali:2025bub}, we study this limit by making use of the following asymptotics for ratios of $q$-Pochhammer symbols:
\begin{align}
\label{eq:qPochhLimit}
   &\frac{\left(x^{2a};x^2\right)_\infty}{\left(x^{2-2a};x^2\right)_\infty}\xrightarrow[]{\beta\to0}\beta^{1-2a}\frac{\Gamma\left(1-a\right)}{\Gamma\left(a\right)}\left(1+o\left(\beta\right)\right)\,,\nn\\
   &\left(Zx^{2a};x^2\right)\xrightarrow[]{\beta\to0}\left(1-Z\right)^{\frac{1}{2}-a}\exp\left({-\frac{\mathrm{Li}_2\left(Z\right)}{\beta}}\right)\left(1+o\left(\beta\right)\right)\,.
\end{align}

It is easier to analyze the result of the limit on the XYZ side
\begin{equation}
\label{eq:XYZLimit1}
   \mathcal{I}_{\text{B}}\xrightarrow[]{\beta\rightarrow 0} \beta^{-1}\frac{\Gamma(\frac{1}{2}+2\phi)\Gamma(\frac{1}{4}\pm \frac{n}{2}-\phi\pm\tau)}{\Gamma(\frac{1}{2}-2\phi)\Gamma(\frac{3}{4}\pm \frac{n}{2}+\phi\mp\tau)}\,.
\end{equation}
In the above, we have dropped subleading terms in the asymptotic expansion. 

On the other hand, the expression for the index of SQED involves a sum over the monopole flux $m$ as well as an integral over the fugacity $u$. Different regions of summation-integration contribute differently to the asymptotics, but the dominant contribution is given as follows \cite{Pasquetti:2019uop,ArabiArdehali:2025bub}. We first change the sum-integral into a double integral, by utilizing the Dirac comb
\begin{equation}
\label{eq:sumintegral}
    \sum_{m\in\mathbb{Z}}\oint \frac{\mathrm{d}u}{2\pi i u}=\sum_{p\in\mathbb{Z}}\int_{-\infty}^\infty \mathrm{d}m \oint \frac{\mathrm{d}u}{2\pi i u} \mathrm{e}^{2\pi i p m}\,.
\end{equation}
As we shall see momentarily, the summation over $p$ will turn out not to be important for our discussion, as only $p=0$ will contribute, but it is generally crucial for studying asymptotics of indices, for example on the second sheet. We choose the scaling for the magnetic flux and the gauge fugacity given by the following change of variables:
\begin{equation}
    \label{eq:holomorphic}
    Z=\mathrm{e}^z=ux^{-m}=u\mathrm{e}^{\frac{\beta m}{2}}\,,\qquad \bar{Z}=\mathrm{e}^{\bar{z}}=u^{-1}x^{-m}=u^{-1}\mathrm{e}^{\frac{\beta m}{2}} \,.
\end{equation}
Essentially, we are thinking of $u$ as the phase of $Z$ and of $x^{-m}$ as its absolute value. Therefore, the integration measure becomes
\begin{equation}
\label{eq:measure}
    \int_{-\infty}^\infty \mathrm{d}m\oint \frac{\mathrm{d}u}{2\pi iu}=\int_\mathbb{C}\frac{\mathrm{d}^2Z}{\pi\beta|Z|^2}\,.
\end{equation}
Using \eqref{eq:qPochhLimit} the integrand can be written as
\begin{align}
\begin{split}
    \mathrm{e}^{2\pi i p m}\left(-u\xi\right)^{m+n}  \frac{(x^{\frac{3}{2}}(x^{-m}u^{-1})^{\pm1}s^{-1};x^2)_{\infty}}{(x^{\frac{1}{2}}(x^{-m}u)^{\pm1}s;x^2)_{\infty}}&=\left(-1\right)^{\frac{z+\bar{z}}{\beta}+n} Z^{-\tau+\frac{n}{2}}\bar{Z}^{-\tau-\frac{n}{2}}\mathrm{e}^{2\pi i p m+\frac{z^2-\bar{z}^2}{2\beta}-\tau \beta n}\\
    &\times\frac{\left(\bar{Z}^{\pm}x^{\frac{3}{2}-2\phi};x^2\right)_{\infty}}{\left(Z^{\pm}x^{2\phi+\frac{1}{2}};x^2\right)_{\infty}}\\
    &\xrightarrow[]{\beta\to0}\frac{(-1)^n}{Z^{\tau-\frac{n}{2}}\bar{Z}^{\tau+\frac{n}{2}}\left|1-Z^\pm\right|^{\frac{1}{2}-2\phi}}\mathrm{e}^{\frac{\mathcal{W}\left(z\right)-\overline{\mathcal{W}}\left(\bar{z}\right)}{\beta}}\,,
    \end{split}
\end{align}
where we have defined the 2d twisted chiral superpotential
\begin{equation}
\label{eq:SQEDTwistedSuperpotential}
    \mathcal{W}\left(z\right)=\frac{z^2}{2}+\mathrm{Li}_2\left(e^z\right)+\mathrm{Li}_2\left(e^{-z}\right)+i\pi z\left(1+2p\right)\,.
\end{equation}
Making use of the identity
\begin{equation}
    \label{eq:dilogarithmreflection}
    \mathrm{Li}_2\left(\mathrm{e}^z\right)+\mathrm{Li}_2\left(\mathrm{e}^{-z}\right)=-\frac{\pi^2}{6}-\frac{\left(\mathrm{Log}\left(-\mathrm{e}^z\right)\right)^2}{2}\,,
\end{equation}
we obtain
\begin{equation}
\mathcal{W}\left(z\right)-\overline{\mathcal{W}}\left(\bar{z}\right)=i\pi \left(1+2p\pm1\right) \left(z+\bar{z}\right)\,,
\end{equation}
where the $\pm1$ depends on the choice of branch for the logarithm. In the limit $\beta\to0$ we can evaluate the integral over $Z$ in the saddle point approximation, however since the twisted superpotential is linear there is no saddle point and the only contribution comes from the term in the sum over $p$ for which $\mathcal{W}\left(z\right)-\overline{\mathcal{W}}\left(\bar{z}\right)$ vanishes. This allows us to eliminate the sum over $p$ as mentioned before, and just end up with
\begin{equation}
    \mathcal{I}_{\text{A}} \xrightarrow[]{\beta\to0}\frac{(-1)^n}{\beta \pi}  \int_\mathbb{C}  \frac{\mathrm{d}^2Z}{Z^{\frac{3}{4}-\frac{n}{2}+\phi+\tau}\bar{Z}^{\frac{3}{4}+\frac{n}{2}+\phi+\tau}\left|1-Z\right|^{1-4\phi}}\,.
\end{equation}

Overall, we find that the 3d index identity \eqref{eq:U1Indexk1} reduces in the $\beta\to0$ Limit 1 to
\begin{align}
    \label{eq:Limit1k1}
        \frac{(-1)^n}{\pi}  \int_{\mathbb{C}}  \frac{\mathrm{d}^2Z}{Z^{\frac{3}{4}-\frac{n}{2}+\phi+\tau}\bar{Z}^{\frac{3}{4}+\frac{n}{2}+\phi+\tau}\left|1-Z\right|^{1-4\phi}}=\frac{\Gamma(\frac{1}{2}+2\phi)\Gamma(\frac{1}{4}\pm \frac{n}{2}-\phi\pm\tau)}{\Gamma(\frac{1}{2}-2\phi)\Gamma(\frac{3}{4}\pm \frac{n}{2}+\phi\mp\tau)}\,.
\end{align}
Notice in particular that all $\beta$ prefactors dropped out and we obtained a finite identity. 

This equality can also be directly proved by noting that the integral is a Coulomb gas integral of the type appearing in correlators of Liouville CFT \cite{Goulian:1990qr,Fateev:2007ab} (we will comment more on this in Section \ref{SubSection:Limit1k1}), which can be evaluated using the contour deformations of Dotsenko and Fateev \cite{Dotsenko:1984ad}, generalized to account for integer mismatches between powers of $Z$ and $\bar{Z}$. In fact, in this way one can prove the more general identity
\begin{align}
\begin{split}
\label{eq:ModifiedDFIntegrals}
    &F\left(a,b,c,d\right)= \frac{1}{\pi} \int_{\mathbb{C}} d^2z \, z^a\left(1-z\right)^c \bar{z}^b \left(1-\bar{z}\right)^d=\frac{ \Gamma\left(1+b\right)\Gamma\left(1+d\right)\Gamma\left(-1-a-c\right)}{\Gamma\left(2+b+d\right)\Gamma\left(-a\right)\Gamma\left(-c\right)}\,,\\
\end{split}
\end{align}
For $a=-(\frac{3}{4}-\frac{n}{2}+\tau+\phi)$, $b=-(\frac{3}{4}+\frac{n}{2}+\tau+\phi)$, $c=d=-\frac{1}{2}+2\phi$ we recover the identity \eqref{eq:Limit1k1}. We provide a proof of this more general identity in Appendix \ref{app:proofDF}.

This identity has already appeared in the context of string amplitudes, see e.g.~\cite{Vanhove:2018elu,Vanhove:2020qtt}. In particular, \eqref{eq:Limit1k1} for $n=0$ is nothing but the Virasoro--Shapiro amplitude for the scattering of four tachyons in flat space \cite{Virasoro:1969me,Shapiro:1970gy}. From this perspective, the index identity \eqref{eq:U1Indexk1} for the abelian mirror duality gives a $q$-deformation of the flat-space Virasoro--Shapiro amplitude.

\subsubsection*{Physical interpretation 1: Mirror symmetry}

The identity \eqref{eq:Limit1k1} has a natural interpretation as an equality of the $S^2$ partition functions of mirror dual 2d $\mathcal{N}=(2,2)$ theories. We start from the following basic mirror pair \cite{Hori:2000kt}:
\begin{itemize}
    \item \textbf{Theory A:} Gauged linear sigma model (GLSM) with $U(1)$ gauge group, a pair of chirals $\Phi$, $\tilde{\Phi}$ of charge $\pm 1$ and no superpotential.
    \item \textbf{Theory B:} Landau--Ginzburg (LG) model of a twisted chiral multiplet $\Sigma$, a pair of twisted chiral multiplets $Y_1$, $Y_2$ with $2\pi $-periodic imaginary part and a twisted superpotential
    \begin{equation}
    \widetilde{\mathcal{W}}=-\frac{1}{4\pi}\left[\Sigma(Y_1-Y_2)+i\mu (\mathrm{e}^{-Y_1}+\mathrm{e}^{-Y_2})\right]+(M_1Y_1+M_2Y_2)\,.
    \end{equation}
\end{itemize}
In the twisted superpotential, $\mu$ is an infrared scale and $M_\alpha=\tilde{m}_\alpha+i\frac{r_\alpha}{4\pi}$, where $\tilde{m}_\alpha$ and $r_\alpha$ are the twisted masses and the R-charges of the chirals $\Phi$, $\tilde{\Phi}$ of the GLSM. These are subject to the relations $\tilde{m}_1-\tilde{m}_2=r_1-r_2=0$ due to the gauge symmetry, and we choose them as\footnote{As commented in the Introduction, more general values of the parameters can be reached by analytic continuation. In particular, different R-charges can be achieved by giving a real part to $\phi$.}
\begin{equation}
    \tilde{m}_1=\tilde{m}_2=-\frac{i}{2\pi}\phi\,,\qquad r_1=r_2=\frac{1}{2}\,.
\end{equation}

As discussed in \cite{Aharony:2017adm} and as we will see in Section \ref{SubSection:Limit2k1}, the associated $S^2$ partition function identity is related to the Limit 2 of the 3d mirror index identity \eqref{eq:U1Indexk1}. Instead, now we shall describe how a different variant of this duality obtained by gauging is related to the Limit 1 identity \eqref{eq:Limit1k1}.

For this purpose, on the side of Theory A we introduce a complexified FI parameter
\begin{equation}
    t=\frac{\theta}{2\pi}+i\eta\,,
\end{equation}
where $\eta$ is the FI parameter and $\theta$ the theta angle. The $S^2$ partition function then reads \cite{Benini:2012ui,Doroud:2012xw} (see Appendix \ref{app:SUSYpf} for our conventions)
\begin{equation}
\label{eq:2dSQEDk1}
\mathcal{Z}_{\text{A}}=\pi^2\sum_{m\in\mathbb{Z}} \int^{\infty}_{-\infty} \mathrm{d}\sigma \, \mathrm{e}^{i m\theta-8\pi^2 i\eta\sigma}  \frac{\Gamma\left(\frac{1}{4}-\phi\pm\left(\frac{m}{2}-2\pi i\sigma\right)\right)}{\Gamma\left(\frac{3}{4}+\phi\pm\left(\frac{m}{2}+2\pi i\sigma\right)\right)}\,.
\end{equation}
On the side of Theory B, this modification corresponds to coupling to a background twisted chiral multiplet $T$ whose scalar component is $t$
\begin{equation}
    \widetilde{\mathcal{W}}=-\frac{1}{4\pi}\left[\Sigma(Y_1-Y_2-2\pi iT)+i\mu (\mathrm{e}^{-Y_1}+\mathrm{e}^{-Y_2})\right]+(M_1Y_1+M_2Y_2)\,.
\end{equation}
The $S^2$ partition function of Theory B, which was shown to be equal to that of Theory A in \cite{Gomis:2012wy}, is
\begin{equation}
    \mathcal{Z}_{\text{B}}
    =\sum_{m\in\mathbb{Z}}\int_{-\infty}^\infty \mathrm{d}\sigma \left[\prod_{\alpha=1,2}\int_{-\infty}^\infty \mathrm{d}x_\alpha\int_{-\pi}^{\pi}\mathrm{d}y_\alpha\right] \mathrm{e}^{4\pi i\sigma(x_1-x_2-2\pi\eta)+im(y_1-y_2 +\theta)+\sum_{\alpha=1,2}\left[2i\mathrm{e}^{-x_\alpha }\sin(y_\alpha)+(2 \phi -\frac{1}{2})x_\alpha\right]}\,.
\end{equation}

In a two-dimensional theory, the theta angle can be viewed as a 0-form background field for a $U(1)$ $(-1)$-form symmetry \cite{Gaiotto:2014kfa}. In fact, the field strength $F$ can be used to construct a current $\star F$ which is a 0-form, so it generates a $(-1)$-form symmetry, and that is conserved by virtue of the Bianchi identity $\mathrm{d}F=0$. The coupling of this current to the background field for the corresponding $(-1)$-form symmetry is identical to that of $\theta$ to the field strength $F$. This symmetry is sometimes also called ``magnetic symmetry" and it exists in generic spacetime dimension $d$ as a $(d-3)$-form symmetry. For 2d (2,2) theories, the multiplet $T$ that contains the complexified FI parameter is the natural supersymmetric completion of the background field for the $(-1)$-form symmetry. 

The variant of the 2d mirror duality that is related to the identity \eqref{eq:Limit1k1} is obtained by gauging such a $(-1)$-form symmetry. We denote the resulting mirror dual theories by A' and B'. At the level of the $S^2$ partition function, this operation amounts to integrating over $\theta\in[-\pi,\pi)$ and $\eta\in\mathbb{R}$. On the side of Theory A, this looks as follows:
\begin{equation}\label{eq:fourier}
   \mathcal{Z}_{\text{A'}}= \int_{-\pi}^\pi \mathrm{d}\theta\int _{-\infty}^{\infty} \mathrm{d}\eta\, \mathrm{e}^{-i\theta n+4\pi i \eta \varphi}\mathcal{Z}_{\text{A}}=\frac{\pi^2}{2}\frac{\Gamma\left(\frac{1}{4}-\phi\pm\left(\frac{n}{2}- i\varphi\right)\right)}{\Gamma\left(\frac{3}{4}+\phi\pm\left(\frac{n}{2}+i\varphi\right)\right)}\,,
\end{equation}
where we introduced the variables $n\in\mathbb{Z}$ and $\varphi\in\mathbb{R}$ that naturally couple to $\theta$ and $\eta$ respectively. The gauging thus corresponds to a Fourier transform, with the new variables $n$, $\varphi$ being the Fourier modes conjugate to $\theta$, $\xi$. In fact, inspired by the similar discussion in 3d due to Kapustin--Strassler \cite{Kapustin:1999ha} that we reviewed in Section \ref{sec:3d}, we can view the gauging operation at the level of the full quantum field theory as a functional Fourier transform of the path integral, where we introduce a 2d BF coupling between the dynamical field $T$ for the gauged $(-1)$-form symmetry and the background vector multiplet $V_b$ with field strength $\Sigma_b$ for a new 0-form global symmetry
\begin{equation}
\mathcal{L}_{\text{BF}}=\int \mathrm{d}^2\tilde{\theta}\,\left[T\,\Sigma_b+h.c.\right]\,.
\end{equation}
In the $S^2$ partition function \eqref{eq:fourier}, the parameters $n$ and $\varphi$ are the flux through $S^2$ and the scalar component for the background vector multiplet $V_b$. 

From \eqref{eq:fourier} we see that the result of the gauging is just the theory of two chirals and that the 0-form symmetry with background $V_b$ is a $U(1)$ symmetry acting on them with opposite charge. This symmetry was gauged in the original Theory A and it became a flavor symmetry in Theory A' after gauging the $(-1)$-form symmetry. This is also analogous to the 3d discussion in Section \ref{sec:3d}, where if we have a $U(1)$ gauge theory, this possesses a $U(1)$ 0-form magnetic symmetry and gauging it freezes the original gauge symmetry to a $U(1)$ 0-form flavor symmetry. We then see that there is a complete analogy between two and three dimensions, and that the gauging of the $(-1)$-form symmetry corresponds to a 2d version of the 3d functional Fourier transform of Kapustin--Strassler. However, unlike the 3d case where the basic $k=1$ duality was invariant under the gauging process, in 2d we get a different looking duality. At the level of the partition functions, the gauging makes us go from the Limit 1 identity \eqref{eq:Limit1k1} to the one we will get in Limit 2 in Section \ref{SubSection:Limit2k1}.

In Section \ref{sec:2dKS} we will further explore this two-dimensional version of the three-dimensional Kapustin--Strassler procedure. We also point out that after gauging the $(-1)$-form symmetry we should get a dual 1-form symmetry, however this turns out to act trivially in Theory A'. Again this is in complete analogy with the fact that in 3d gauging the $U(1)$ 0-form magnetic symmetry of SQED results in a trivially acting dual 1-form symmetry.

Let us see the analogue in Theory B of the gauging of the $(-1)$-form symmetry and how this allows us to connect to the Limit 1 identity \eqref{eq:Limit1k1}
\begin{align}
\begin{split}
\label{eq:gaugedtopB}
   \mathcal{Z}_{\text{B'}}=&\int_{-\pi}^\pi \mathrm{d}\theta\int _{-\infty}^{\infty} \mathrm{d}\eta\quad \mathrm{e}^{-i\theta n+4\pi i \eta \varphi}\mathcal{Z}_{\text{B}}\\
   &=\int_{-\pi}^\pi \mathrm{d}\theta\int _{-\infty}^{\infty} \mathrm{d}\eta\sum_{m\in\mathbb{Z}}\int_{-\infty}^\infty \mathrm{d}\sigma \left[\prod_{\alpha=1,2}\int_{-\infty}^\infty \mathrm{d}x_\alpha\int_{-\pi}^{\pi}\mathrm{d}y_\alpha\right] \\
   &\times\mathrm{e}^{-i\theta n+4\pi i \eta \varphi+4\pi i\sigma(x_1-x_2-2\pi\eta)+im(y_1-y_2 +\theta)+2i\mathrm{e}^{-x_1 }\sin(y_1)+2i\mathrm{e}^{-x_2 }\sin(y_2)+(2 \phi -\frac{3}{2})(x_1+x_2)}\,.
   \end{split}
\end{align}
We now change variables
\begin{equation}
    x_\pm=\frac{x_1\pm x_2}{2}\,,\qquad y_\pm=\frac{y_1\pm y_2}{2}
\end{equation}
and integrate over $y_+$, $x_+$ using that
\begin{align}\label{eq:Bessel}
    &\int ^\pi_{-\pi}\mathrm{d}y_+\, \mathrm{e}^{ia\sin(y_+)}=2\pi J_0(a)\,,\nn\\
    &\int_{-\infty}^\infty \mathrm{d}x_+\, \mathrm{e}^{-qx_+} J_0(2b\mathrm{e}^{-x_+})=\frac{\Gamma\left(\frac{q
    }{2}\right)}{2b^q\Gamma\left(1-\frac{q}{2}\right)}\,\quad\mathrm{
    for
    }\quad \mathrm{Re}(q)>0\,,
\end{align}
where $J_0$ is the Bessel function of the first kind of order zero. Further performing the change of variables
\begin{equation}
    z=\mathrm{e}^{2(x_--iy_-)+i\pi}
\end{equation}
results in
\begin{equation}
   \mathcal{Z}_{\text{B'}} =\pi \frac{\Gamma\left(\frac{1}{2}-2\phi\right)}{\Gamma(\frac{1}{2}+2\phi)}\int_{-\pi}^\pi \mathrm{d}\theta\int _{-\infty}^{\infty} \mathrm{d}\eta\sum_{m\in\mathbb{Z}}(-1)^m\int_{-\infty}^\infty \mathrm{d}\sigma\int d^2z \frac{\mathrm{e}^{-i\theta n+4\pi i \eta \varphi-8\pi^2 i\sigma\eta+im\theta}}{|z|^{\frac{3}{2}+2\phi-4\pi i\sigma}|1-z|^{1-4\phi}}\left(\frac{\bar{z}}{z}\right)^{\frac{m}{2}}\,.
\end{equation}

The original sum over $m$ and integral over $\sigma$ can now be evaluated yielding delta functions, which we can use to perform the integral over $z$
\begin{equation}
   \mathcal{Z}_{\text{B'}}= \frac{\pi}{2}(-1)^n\frac{\Gamma(\frac{1}{2}-2\phi)}{\Gamma(\frac{1}{2}+2\phi)}\int_\mathbb{C} \mathrm{d}^2Z \frac{1}{|Z|^{\frac{3}{2}+2\phi-2i\varphi}|1-Z|^{1-4\phi}}\left(\frac{Z}{\bar{Z}}\right)^\frac{n}{2}\,,
\end{equation}
where we have defined $Z=\mathrm{e}^{2\pi\eta-i\theta}$.  We finally see that the identity \eqref{eq:Limit1k1} corresponds to the equality of the partition functions of Theory A' and B' (with the identification of parameters $\varphi=i\tau$)
\begin{align}
    \mathcal{Z}_{\text{B'}}&=\frac{\pi}{2}(-1)^n  \frac{\Gamma(\frac{1}{2}-2\phi)}{\Gamma(\frac{1}{2}+2\phi)}\int_\mathbb{C} \mathrm{d}^2Z \frac{1}{|Z|^{\frac{3}{2}+2\phi-2i\varphi}|1-Z|^{1-4\phi}}\left(\frac{Z}{\bar{Z}}\right)^\frac{n}{2}\nn\\
    &=\frac{\pi^2}{2}\frac{\Gamma\left(\frac{1}{4}-\phi\pm\left(\frac{n}{2}- i\varphi\right)\right)}{\Gamma\left(\frac{3}{4}+\phi\pm\left(\frac{n}{2}+ i\varphi\right)\right)}=\mathcal{Z}_{\text{A'}}\,.
\end{align}
We point out in particular that the extra refinement of the 3d index by the background flux $n$ for the topological symmetry of SQED resulted in 2d in a background flux for the $U(1)$ flavor symmetry that we gained in the process of gauging the $(-1)$-form symmetry.

\subsubsection*{Physical interpretation 2: Liouville CFT}

The second physical interpretation of the Limit 1 identity \eqref{eq:Limit1k1} is in the context of 2d Liouville CFT. This has already been observed in the literature, see e.g.~\cite{Aganagic:2013tta,Aganagic:2014oia,Pasquetti:2019uop}. Here we will just briefly review this fact and in the later sections we will see some generalizations.

Liouville theory is a CFT of a single scalar field $\phi(z,\bar{z})$ with an exponential potential
\begin{equation}
    \mathcal{L}=\frac{1}{4\pi}(\partial_m\phi)^2+\mu\mathrm{e}^{2b\phi} \,,
\end{equation}
where $\mu$ is the cosmological constant and $b$ the Liouville coupling. It has central charge
\begin{equation}
    c=1+6Q^2\,,\qquad Q=b+b^{-1}
\end{equation}
and the basic objects are the vertex operators
\begin{equation}
    V_{\alpha}(z,\bar{z})=\mathrm{e}^{2\alpha\phi(z,\bar{z})}
\end{equation}
with dimensions
\begin{equation}
    \Delta(\alpha)=\alpha(Q-\alpha)\,.
\end{equation}

By first integrating over the zero modes $\phi_0$ of the boson $\phi(z,\bar{z})=\phi_0+\tilde{\phi}(z,\bar{z})$, it was noted in \cite{Goulian:1990qr,Fateev:2007qn,Fateev:2007ab} that the residue of the correlation function of vertex operators can be expressed as the correlator in a free theory, which in turn admits a Coulomb gas or Dotsenko--Fateev integral representation \cite{Dotsenko:1984ad}
\begin{equation}
\label{eq:LiouvilleCorr}
    \mathrm{Res}_{\alpha=Q-Nb} \langle V_{\alpha_1}(y_1,\bar{y}_1)\dots V_{\alpha_M}(y_M,\bar{y}_M)\rangle=(\pi \mu)^n \prod_{i<j}^M \left|y_i-y_j\right|^{-4\alpha_i\alpha_j} \int \mathfrak{D}_N(z)\prod_{a=1}^N\prod_{i=1}^M\left|z_a-y_i\right|^{-4b\alpha_i} \,.
\end{equation}
In the above $\alpha_i$ are the momenta of the vertex operators, $\alpha=\sum_{i=1}^M \alpha_i$ is the total momentum which has to satisfy the neutrality coming from the residue
\begin{equation}
    \alpha=\sum_{i=1}^M \alpha_i=Q-Nb\,,
\end{equation}
and the measure is given by the Vandermonde determinant
\begin{equation}
    \mathfrak{D}_N(z)=\frac{1}{\pi^N N!}\prod_{a<b}^n\left|z_a-z_b\right|^{-4b^2} \prod_{a=1}^N d^2z_a\,.
\end{equation}

It is immediate to see that the integral on the l.h.s.~of \eqref{eq:Limit1k1} in the case in which the parameter $n$ is turned off $n=0$ has exactly the form of one of the Coulomb gas integrals for the Liouville CFT. In particular, we have the one-dimensional $N=1$ Coulomb gas integral for the three-point function, after using the $SL(2,\mathbb{C)}$ invariance to fix the positions of the operators to $(z_1,z_2,z_3)=(0,1,\infty)$. The momenta are identified as in \cite{Pasquetti:2019uop}\footnote{For Abelian gauge groups, we are unable to determine the value of $b$, which for non-Abelian gauge theories corresponding to higher-dimensional Coulomb gas integrals can be read off from $\mathfrak{D}_N(z)$.}
\begin{equation}
    4b\alpha_1=\frac{3}{2}+2\tau+2\phi\,,\qquad 4b\alpha_2=1-4\phi\,,
\end{equation}
while $b\alpha_3$ is determined by the neutrality condition
\begin{equation}
    b(\alpha_1+\alpha_2+\alpha_3)=1\quad\Rightarrow\quad 4b\alpha_3=\frac{3}{2}-2\tau+2\phi\,.
\end{equation}
The identity \eqref{eq:Limit1k1} coincides with the $N=1$ residue of the DOZZ formula for the three-point function of Liouville CFT \cite{Dorn:1994xn,Zamolodchikov:1995aa}.

\subsection{Limit 2 of the index}
\label{SubSection:Limit2k1}

The second possible limit of the 3d index identity \eqref{eq:U1Indexk1}, which we shall refer to as Limit 2, is
\begin{equation}\label{eq:Limit2scaling}
    \textbf{Limit 2}:\qquad s=x^{2\phi}\,,\qquad \xi,\phi\,\text{ finite}\,.
\end{equation}

On the SQED side, we also take
\begin{equation}\label{eq:Limit2k1gauge}
    u=\mathrm{e}^{2\pi i \beta \sigma}\,.
\end{equation}
Using \eqref{eq:qPochhLimit}, we get
\begin{align}\label{eq:k=1_SQED_2d}
    \mathcal{I}_{\text{A}}\xrightarrow[]{\beta \rightarrow0} \beta^{4\phi}\sum_{m\in \mathbb{Z}}\int_{-\infty}^{\infty}\mathrm{d}\sigma (-\xi)^{m+n} \frac{\Gamma (\frac{1}{4}+\phi\pm (\frac{m}{2}+2\pi i \sigma))}{\Gamma (\frac{3}{4}-\phi\pm (\frac{m}{2}-2\pi i \sigma))}\,.
\end{align}

On the XYZ side we instead find
\begin{align}\label{eq:k=1_XYZ_2d}
    \mathcal{I}_{\text{B}}\xrightarrow[]{\beta \rightarrow0} \beta^{4\phi}(1-\xi)^{-\frac{1}{2}-2\phi+n}(1-\xi^{-1})^{-\frac{1}{2}-2\phi-n}\frac{\Gamma(\frac{1}{2}+2\phi)}{\Gamma (\frac{1}{2}-2\phi)}\,.
\end{align}

Equating \eqref{eq:k=1_SQED_2d} and \eqref{eq:k=1_XYZ_2d} we find that the $\beta$ prefactors cancel out and we get that the 3d index identity \eqref{eq:U1Indexk1} reduces in the $\beta\to0$ Limit 2 to
\begin{equation}\label{eq:Limit2k1}
    \sum_{m\in \mathbb{Z}}\int_{-\infty}^{\infty}\mathrm{d}\sigma (-\xi)^{m} \frac{\Gamma (\frac{1}{4}+\phi\pm (\frac{m}{2}+2\pi i \sigma))}{\Gamma (\frac{3}{4}-\phi\pm (\frac{m}{2}-2\pi i \sigma))}=(1-\xi)^{-\frac{1}{2}-2\phi}(1-\xi^{-1})^{-\frac{1}{2}-2\phi}\frac{\Gamma(\frac{1}{2}+2\phi)}{\Gamma (\frac{1}{2}-2\phi)}\,.
\end{equation}
Note that also the dependence on $n$ dropped out, since it was just in an overall prefactor that was equal on the two sides.

This equality can also be proved in two different ways. One consists of explicitly performing the contour integration and picking up suitable residues. Details of this computation are provided in Appendix \ref{app:Limit2k1residueproof}. The second proof instead uses the Limit 1 identity \eqref{eq:Limit1k1} and thus shows that the Limit 2 identity \eqref{eq:Limit2k1} is not independent. Indeed, starting from the l.h.s.~of \eqref{eq:Limit2k1}, we can first use \eqref{eq:ModifiedDFIntegrals} to replace the ratio of Gamma functions with an auxiliary Coulomb gas-like integral, taking 
\begin{align}
    a_{\pm}&=-\frac{3}{4}+\phi \mp \left(\frac{m}{2}+2\pi i\sigma\right)\,,\qquad c=-\frac{1}{2} -2\phi\,,\\
    b_{\pm}&=-\frac{3}{4}+\phi \pm\left(\frac{m}{2} -2\pi i\sigma\right)\,,\qquad d=-\frac{1}{2} -2\phi\,,
\end{align}
where $\{a_+,b_+,c,d\}$ correspond to the case where $m\geq 0$ and $\{a_-,b_-,c,d\}$ for $m < 0$. At this point, we first integrate over $\sigma$ and sum over $m$ so to get delta-functions, which we can then use to evaluate the auxiliary Coulomb gas-like integral using again \eqref{eq:ModifiedDFIntegrals} to get precisely the r.h.s.~of \eqref{eq:Limit2k1}
\begin{equation}
\begin{split}
    \text{l.h.s.~of \eqref{eq:Limit2k1}}&=\frac{\Gamma(\frac{1}{2}+2\phi)}{\Gamma (\frac{1}{2}-2\phi)}\sum_{m\in\mathbb{Z}}\int_{-\infty}^{\infty}\mathrm{d}\sigma \; \xi^{m}\left( F(a_+,b_+,c,d)H(m)+ F(a_-,b_-,c,d)H(-m-1)\right)\\
    &=\frac{\Gamma(\frac{1}{2}+2\phi)}{\pi\Gamma (\frac{1}{2}-2\phi)}\sum_{m\in \mathbb{Z}}\int_{-\infty}^{\infty}\mathrm{d}\sigma \; (-\xi)^{m}(-1)^m \int_{\mathbb{C}} \mathrm{d}^2z\, |1-z|^{c+d} \left(\frac{1-\bar{z}}{1-z}\right)^{\frac{d-c}{2}}\\&\times \left(|z|^{a_++b_+}\left(\frac{\bar{z}}{z}\right)^{\frac{b_+-a_+}{2}}H(m)+|z|^{a_-+b_-}\left(\frac{\bar{z}}{z}\right)^{\frac{b_--a_-}{2}}H(-m-1)\right)\\
    &=(1-\xi)^{-\frac{1}{2}-2\phi}(1-\xi^{-1})^{-\frac{1}{2}-2\phi}\frac{\Gamma(\frac{1}{2}+2\phi)}{\Gamma (\frac{1}{2}-2\phi)}=\text{r.h.s.~of \eqref{eq:Limit2k1}}\,,
\end{split}
\end{equation}
where $H(x)$ is the Heaviside step function. This derivation is reminiscent of how we showed in Section \ref{sec:3d} that in 3d the basic $k=1$ duality is left invariant by the gauging of the $U(1)$ topological symmetry. In 2d instead we get two different looking identities, which however are not independent of each other.

\subsubsection*{Physical interpretation: Mirror symmetry}

As we already mentioned in Section \ref{SubSection:Limit1k1}, the Limit 2 identity \eqref{eq:Limit2k1} has a direct interpretation in terms of a 2d mirror duality between a GLSM with $U(1)$ gauge group and a pair of chirals of charge $\pm1$, and a LG model of twisted chiral fields \cite{Hori:2000kt}. 

The $S^2$ partition function of the GLSM for generic values of the gauge and R-charges, and twisted masses is given by \cite{Benini:2012ui,Doroud:2012xw}
\begin{align}\label{eq:GLSM_k=1}
    \mathcal{Z}_{\text{A}}=\pi^2\sum_{m\in\mathbb{Z}}\mathrm{e}^{im\theta}\int_{-\infty}^\infty \mathrm{d}\sigma \; \mathrm{e}^{-8\pi^2 i  \sigma \eta} \prod_{\alpha=1,2}\frac{\Gamma (\frac{r_\alpha}{2}-2\pi i\tilde{m}_\alpha-\frac{mQ_\alpha}{2}-2\pi i  Q_\alpha\sigma)}{\Gamma (1-\frac{r_\alpha}{2}+2\pi i\tilde{m}_\alpha-\frac{mQ_\alpha}{2}+2\pi i Q_\alpha\sigma)}\,.
\end{align}
For us $Q_1=1$ and $Q_2=-1$. By identifying
\begin{equation}\label{eq:parid}
    \mathrm{e}^{i\theta}=-\xi\,,\qquad \eta=0\,,\qquad \tilde{m}_1=\tilde{m}_2=i\frac{\phi}{2\pi}\,,\qquad r_1=r_2=\frac{1}{2}\,,
\end{equation}
this gives the l.h.s.~of the Limit 2 identity \eqref{eq:Limit2k1}.

In \cite{Gomis:2012wy} it was shown that the partition function of the GLSM is equal to that of the mirror LG model
\begin{equation}\label{eq:LGk1}             
    \mathcal{Z}_{\text{B}}=\sum_{m\in\mathbb{Z}}\int_{-\infty}^\infty \mathrm{d}\sigma \prod_{\alpha=1,2}\left[\int_{-\infty}^\infty \mathrm{d}x^\alpha\int_{-\pi}^{\pi}\mathrm{d}y^\alpha \right]\mathrm{e}^{4\pi i\sigma(Q_\alpha x^\alpha-2\pi \eta)+im(Q_\alpha y^\alpha+\theta)+2ie^{-x^\alpha }\sin(y^\alpha)+(4\pi i \tilde{m}_\alpha -r_\alpha) x^\alpha}\,,
\end{equation}
We would like to show that this is equivalent to the r.h.s.~of \eqref{eq:Limit2k1}, under the same identification of parameters \eqref{eq:parid}. 

For this, we consider the change of variables
\begin{equation}
    x_\pm=\frac{x_1\pm x_2}{2}\,,\qquad y_\pm=\frac{y_1\pm y_2}{2}
\end{equation}
Then, performing the integral over $x^+$, $y^+$ using \eqref{eq:Bessel} yields
\begin{align}
    \mathcal{Z}_{\text{B}}=\pi\frac{\Gamma(\frac{1}{2}+2\phi)}{\Gamma (\frac{1}{2}-2\phi)}\sum_{m\in\mathbb{Z}}(-\xi)^m \int_{-\infty}^\infty \mathrm{d}\sigma \int_\mathbb{C} \mathrm{d}^2 z \frac{1}{|z|^{\frac{3}{2}-2\phi+4\pi i \sigma}}\left(\frac{z}{\bar{z}}\right)^{\frac{m}{2}}\frac{1}{|1-z|^{1+4\phi}}
\end{align}
where we defined $z=e^{2(x^--iy^-)+i\pi}$. Now, integrating out the field strength multiplet $\Sigma$ and then performing the $z$-integral yields
\begin{align}
    \mathcal{Z}_{\text{B}}&=\pi\frac{\Gamma(\frac{1}{2}+2\phi)}{\Gamma (\frac{1}{2}-2\phi)}\int_\mathbb{C} \mathrm{d}^2 z \ \frac{\frac{1}{2}\delta(|z|-1)2\pi \delta(\arg(z)+\arg(\xi))}{|z|^{\frac{3}{2}-2\phi}}\frac{1}{|1-z|^{1+4\phi}}\\
    &=\pi^2(1-\xi)^{-\frac{1}{2}-2\phi}(1-\xi^{-1})^{-\frac{1}{2}-2\phi}\frac{\Gamma(\frac{1}{2}+2\phi)}{\Gamma (\frac{1}{2}-2\phi)}.
\end{align}
In this sense, the Limit 2 identity \eqref{eq:Limit2k1} can be understood as a manifestation of 2d mirror symmetry.


\section{Reduction of the duality for SQED with  $k$ flavors}
\label{sec:limitk}

\subsection{Limit 1 of the index}
\label{SubSection:Limit1genk}

The limits that we have studied in the previous section can be generalized to the case of $k$ flavors. Recalling the associated 3d index identity \eqref{eq:U1Index}, this time we have $k$ parameters $s_i$ which are taken to scale in the limit $\beta\to0$ as the parameter $s$ for $k=1$, but we also have the $k-1$ parameters $v_i$ for which we have to prescribe a scaling. The generalization of the Limit 1 \eqref{eq:Limit1scaling} corresponds to taking
\begin{equation}
    \textbf{Limit 1}:\qquad\xi=x^{2\tau}\,,\quad s_i=x^{2\phi_i}\,, \quad v_i,\tau,\phi_i\, \,\text{ finite}\,,\quad i=1,\cdots,k\,.
\end{equation}
We also remind the reader that we have the constraint $\prod_{i=1}^k v_i=1$, and thus only $k-1$ linearly independent $v_i$.

On the side of SQED with $k$ flavors, we proceed as for $k=1$. Namely, we first change variables
\begin{equation}
    Z=\mathrm{e}^z=ux^{-m}=u\mathrm{e}^{\frac{\beta m}{2}}\,,\qquad \bar{Z}=\mathrm{e}^{\bar{z}}=u^{-1}x^{-m}=u^{-1}\mathrm{e}^{\frac{\beta m}{2}} \,.
\end{equation}
Then, we introduce the Dirac comb \eqref{eq:sumintegral} to obtain the measure \eqref{eq:measure}, study the $\beta\to0$ limit using \eqref{eq:qPochhLimit} and argue from the resulting twisted superpotential that only the term $p=0$ survives. We avoid spelling out all the technical details, since the computation is a straightforward generalization of the one for $k=1$. Overall, we find that the first line of \eqref{eq:U1Index} gives in the $\beta\to0$ Limit 1
\begin{equation}
   \mathcal{I}_{\text{A}} \xrightarrow[]{\beta\to0}  \frac{1}{\pi\beta}\int_{\mathbb{C}}\mathrm{d}^2Z \frac{1}{|Z|^{2+2\tau}}\prod_{i=1}^k\left[\frac{1}{|v_iZ|^{-\frac{1}{2}+2\phi_i}|1-Zv_i|^{1-4\phi_i}}\right] \,.
\end{equation}

On the other hand, on the side of Theory B given by the abelian quiver gauge theory, we should probe the integration region given by the following scaling of the gauge fugacities and fluxes:
\begin{equation}
    u_a=\mathrm{e}^{2\pi i\beta\sigma_a}\,,\quad a=1,\cdots,k-1\,.
\end{equation}
Note that this is analogous to what we did in \eqref{eq:Limit2k1gauge} for the Limit 2 of the $k=1$ case. Using again the asymptotics of the $q$-Pochhammer \eqref{eq:qPochhLimit}, we then obtain
\begin{align}
\begin{split}
    \mathcal{I}_{\text{B}} \xrightarrow[]{\beta\to0} & \frac{1}{\beta}\prod_{i=1}^k\left[\frac{\Gamma(\frac{1}{2}+2\phi_i)}{\Gamma(\frac{1}{2}-2\phi_i)}\right] \sum_{\vec{m}\in\mathbb{Z}^{k-1}}\int \prod_{a=1}^{k-1}\left[\mathrm{d}\sigma_a\left(\frac{v_a}{v_{a+1}}\right)^{-m_a}\right]\\
    &\times\prod_{i=1}^{k}\left[\frac{\Gamma(\frac{1}{4}-\phi_i\pm(\tau\delta_{i,k}+\frac{m_i-m_{i-1}}{2}+2\pi i (\sigma_{i-1}-\sigma_i)) }{\Gamma(\frac{3}{4}+\phi_i\pm(\tau\delta_{i,k}+\frac{m_{i-1}-m_{i}}{2}+2\pi i (\sigma_{i-1}-\sigma_i)) }\right]\,.
    \end{split}
\end{align}

As usual, the $\beta$ prefactor cancels between the two sides, resulting in the following finite identity:
\begin{align}
    \begin{split}
        \label{eq:Limit1k}
         &\frac{1}{\pi}\int_{\mathbb{C}}\mathrm{d}^2Z \frac{1}{|Z|^{2+2\tau}}\prod_{i=1}^k\left[\frac{1}{|v_iZ|^{-\frac{1}{2}+2\phi_i}|1-Zv_i|^{1-4\phi_i}}\right]=\prod_{i=1}^k\left[\frac{\Gamma(\frac{1}{2}+2\phi_i)}{\Gamma(\frac{1}{2}-2\phi_i)}\right]\\
         &\times\sum_{\vec{m}\in\mathbb{Z}^{k-1}} \prod_{a=1}^{k-1}\left[\int_{-\infty}^\infty\mathrm{d}\sigma_a\left(\frac{v_a}{v_{a+1}}\right)^{-m_a}\right]\prod_{i=1}^{k}\left[\frac{\Gamma(\frac{1}{4}-\phi_i\pm(\tau\delta_{i,k}+\frac{m_i-m_{i-1}}{2}+2\pi i (\sigma_{i-1}-\sigma_i)) }{\Gamma(\frac{3}{4}+\phi_i\pm(\tau\delta_{i,k}+\frac{m_{i-1}-m_{i}}{2}+2\pi i (\sigma_{i-1}-\sigma_i)) }\right]\,.
    \end{split}
\end{align}
In Section \ref{sec:2dKS} we will provide a proof of this identity assuming only the validity of the Limit 1 identity \eqref{eq:Limit1k1} for $k=1$ using a two-dimensional version of the Kapustin--Strassler procedure.

\subsubsection*{Physical Interpretation 1: Mirror symmetry}

For $k\geq 2$, the identity \eqref{eq:Limit1k} obtained in Limit 1 can be interpreted in terms of the 2d mirror symmetry between the following theories:
\begin{itemize}
    \item \textbf{Theory B:} GLSM with $U(1)^{k-1}$ gauge group and $k$ pairs of chiral fields $\Phi_{2i-1}$, $\Phi_{2i}$ where $i=1,\cdots,k$. The charge of the $j$-th chiral field under the $a$-th $U(1)$ gauge group is labeled by the charge matrix $Q^a_j$, where $a=1,\cdots,k-1$ and $j=1,\cdots,2k$. We specialize here to the following charge matrix:
    \begin{equation}\label{eq:charges}
		Q^a_j= \left(
		\begin{array}{cccccc cccccc}
			1 & -1 & -1 & 1 & 0 & 0 &0 & \cdots & 0 & 0 & 0 & 0 \\
			0 & 0 & 1 & -1 & -1 & 1 &0 & \cdots & 0 & 0 & 0 & 0\\
			 &  &  &\vdots  &  &  &\ddots &  &  & \vdots &  & \\
			0 & 0 & 0 & 0 & 0 & \cdots & & 0 & 1 & -1 & -1 & 1 \\
		\end{array}
		\right)\,.
	\end{equation}
    There is no superpotential for the chirals. 
    \item \textbf{Theory A:} LG model of $k-1$ twisted chiral multiplets $\Sigma_a$, $k$ pairs of twisted chiral multiplets $Y_{2i-1}$, $Y_{2i}$ with $2\pi $-periodic imaginary part and a twisted superpotential
    \begin{equation}
    \widetilde{\mathcal{W}}=-\frac{1}{4\pi}\left[\sum_{a=1}^{k-1}\Sigma_a\left(\sum^{2k}_{j=1}Q^{a}_jY^{j}\right)+i\mu \sum_{a=1}^{2k}\mathrm{e}^{-Y^j}\right]+M_jY^j\,.
    \end{equation}
\end{itemize}
The apparently inverted labelling of these two theories is due to the fact that the GLSM is the 2d version of the same Theory B that we had in 3d, see Section \ref{sec:3d}.

The $S^2$ partition function of the GLSM is given by
\begin{equation} 
    \mathcal{Z}_{\text{B}}=\pi^{2k}\sum_{\vec{m}\in\mathbb{Z}^{k-1}} \mathrm{e}^{im_a\theta^a}\prod_{a=1}^{k-1}\left[\int_{-\infty}^\infty \mathrm{d}\sigma_a\right] e^{-8\pi^2 i \sigma_a \eta^a} \prod_{j=1}^{2k} \frac{\Gamma (\frac{r_j}{2}-2\pi i\tilde{m}_j+\frac{m_aQ^a_j}{2}-2\pi i\sigma_aQ^a_j)}{\Gamma (1-\frac{r_j}{2}+2\pi i\tilde{m}_j+\frac{m_aQ^a_j}{2}+2\pi i \sigma_aQ^a_j)}\,.
\end{equation}
In order to match with the r.h.s.~of \eqref{eq:Limit1k} we should use the charges $Q^q_j$ in \eqref{eq:charges} and identify the parameters as follows:
\begin{align}\label{eq:paridentification}
    &\mathrm{e}^{i\theta^a}=\frac{v_{a+1}}{v_{a}}\,,\quad \eta^a=0\,,\quad a=1,\cdots,k-1\,,\nn\\
    &r_j=\frac{1}{2}\,,\quad j=1,\cdots,2k\,,\nn\\
    &\tilde{m}_j=\begin{cases}
        -\frac{i}{2\pi}(\phi_i-\tau\delta_{i,k}) & j=2i-1\,, \\
        -\frac{i}{2\pi}(\phi_i+\tau\delta_{i,k}) & j=2i\,.
    \end{cases}
\end{align}

On the other hand, the $S^2$ partition function of the mirror dual LG model is given by
\begin{equation}
\begin{split}
    \mathcal{Z}_{\text{A}}=&\sum_{\vec{m}\in\mathbb{Z}^{k-1}}\prod_{a=1}^{k-1}\left[\int_{-\infty}^\infty \mathrm{d}\sigma_a\right]\prod_{j=1}^{2k}\left[ \int_{-\infty}^\infty \mathrm{d} x^j \int_{-\pi}^\pi \mathrm{d} y^j\right] \mathrm{e}^{4\pi i \sigma_a (Q^a_j x^j -2\pi \eta^a)+im_a(Q^a_j y^j +\theta^a)+2i\mathrm{e}^{-x^j}\sin (y^j)+(4\pi i \tilde{m}_j-r_j)x^j}\\
    &=\sum_{\vec{m}\in\mathbb{Z}^{k-1}}\prod_{a=1}^{k-1}\left[\int_{-\infty}^\infty \mathrm{d}\sigma_a\right] \mathrm{e}^{im_a\theta^a-8\pi^2 i \sigma_a \eta^a}\prod_{i=1}^k\prod_{\alpha=1,2}\left[ \int_{-\infty}^\infty \mathrm{d} x^\alpha_i \int_{-\pi}^\pi \mathrm{d} y^\alpha_i\right]\\
    &\times \mathrm{e}^{4\pi i (\sigma_{i} -\sigma_{i-1}) (x^1_i-x^2_i) +i (m_{i} -m_{i-1} )(y^1_i-y^2_i) +2i\mathrm{e}^{-x^\alpha_i}\sin (y^\alpha_i)+(2\phi_i-\frac{3}{2})(x^1_i+x^2_i)-2\tau(x^1_k-x^2_k)}\,.
\end{split}
\end{equation}
We introduced $m_0=m_k=0$ and $\sigma_0=\sigma_k=0$ for ease of notation. Similarly to what we did in the previous sections when dealing with LG partition functions, we now perform a change of variables for each pair of twisted chirals
\begin{equation}
    x^{\pm}_i=\frac{x^1_i\pm x^2_i}{2}\,,\qquad y^{\pm}_i=\frac{y^2_i\pm y^1_i}{2}\,,\qquad i=1,\cdots,k
\end{equation}
and then integrate over all $x^+_i$, $y^+_i$ using \eqref{eq:Bessel}, so to get
\begin{equation}
\begin{split}
    \mathcal{Z}_{\text{A}}
    =&\pi^k\sum_{\vec{m}\in\mathbb{Z}^{k-1}}\prod_{a=1}^{k-1}\left[\int_{-\infty}^\infty\mathrm{d}\sigma_a\left(\frac{v_a}{v_{a+1}}\right)^{-m_a}\right]\\
    &\times\prod_{i=1}^k\left[\frac{\Gamma(\frac{1}{2}-2\phi_i)}{\Gamma (\frac{1}{2}+2\phi_i)}\int_{\mathbb{C}} \mathrm{d}^2 z_i \ \frac{1}{|z_i|^{\frac{3}{2}+2\phi_i-4\pi i (\sigma_i-\sigma_{i-1})+2\tau \delta_{i,k}}}\left(\frac{z_i}{\bar{z}_i}\right)^{\frac{m_i-m_{i-1}}{2}}\frac{1}{|1-z_i|^{1-4\phi_i}}\right]\,,
\end{split}
\end{equation}
where we defined $z_i=\mathrm{e}^{2(x^-_i-iy^-_i)+i\pi}$ and we used the identification of parameters \eqref{eq:paridentification}. Next, we can perform the sum-integral for each field strength multiplet $\Sigma_i$, which gives $2(k-1)$ delta functions for $z_i$. Finally, we can use these to eliminate the integrals over $z_i$ for $i=1,\cdots,k-1$ and redefine $Z= \frac{z_k}{v_k}$ to arrive at Coulomb gas integral on the l.h.s.~of \eqref{eq:Limit1k}
\begin{equation}
\begin{split}
    \mathcal{Z}_{\text{A}}&=\pi^{2k-1}\prod_{i=1}^k\left[\frac{\Gamma(\frac{1}{2}-2\phi_i)}{\Gamma (\frac{1}{2}+2\phi_i)}\int_{\mathbb{C}} \mathrm{d}^2 z_i\right]\frac{\prod_{i=1}^{k-1}\delta(\log(|z_{i+1}|/|z_i|))\delta(\arg(z_{i+1})-\arg(z_i)+\arg(v_{i})-\arg(v_{i+1}))}{\prod_{i=1}^k|z_i|^{\frac{3}{2}+2\phi_i+2\tau \delta_{i,k}}|1-z_i|^{1-4\phi_i}}\\
    &=\pi^{2k-1}\prod_{i=1}^k\left[\frac{\Gamma(\frac{1}{2}-2\phi_i)}{\Gamma (\frac{1}{2}+2\phi_i)}\right]\int_{\mathbb{C}} \mathrm{d}^2 Z\frac{1}{|Z|^{2+2\tau}}\prod_{i=1}^k\left[\frac{1}{|v_iZ|^{-\frac{1}{2}+2\phi_i}|1-Zv_i|^{1-4\phi_i}}\right]\,.
\end{split}
\end{equation}
We have thus shown that the identity between the $S^2$ partition functions of the 2d mirror dual theories is equivalent to the Limit 1 identity in \eqref{eq:Limit1k}.

\subsubsection*{Physical Interpretation 2: Liouville CFT}

In Section \ref{SubSection:Limit1k1} we have seen that the Limit 1 identity for $k=1$ has an interpretation in Liouville CFT. Specifically, the side of the identity that can be related to a LG partition function coincides with the Coulomb gas integral for a certain correlation function of Liouville. The same turns out to be true for the generic $k$ case.

Specifically, the l.h.s.~of \eqref{eq:Limit1k} also takes the form of a Liouville Coulomb gas integral with $N=1$ screening charges, whose generic form we recall was given in \eqref{eq:LiouvilleCorr}. However, this time we have a $(k+2)$-point correlation function. The positions of the operators can be taken to be

\begin{equation}
    (z_1,z_{i+1} , z_{k+2})=(0,v^{-1}_i,\infty)\,,\qquad \prod_{i=1}^kv_i=1\,.
\end{equation}
The momenta are instead identified as\footnote{As for $k=1$, the Liouville coupling $b$ is left undetermined.}
\begin{equation}
    4b\alpha_1=2-\frac{k}{2}+2\tau+2\Phi\,,\quad 4b\alpha_{i+1}=1-4\phi_i\,,\quad 4b\alpha_{k+2}=2-\frac{k}{2}-2\tau+2\Phi\,.
\end{equation}
In the above we have defined $\Phi=\sum_{i=1}^k\phi_i$, and as in the $k=1$ case $b\alpha_{k+2}$ is determined by the neutrality condition.

We can summarize the general dictionary between the parameters of the GLSM and those of the Liouville correlator
\begin{align}
    \text{FIs}\,\,&\longleftrightarrow\,\,\text{positions}\nn\\
    \text{masses}\,\,&\longleftrightarrow\,\,\text{momenta}\,.
\end{align}

\subsection{Limit 2 of the index}
\label{SubSection:Limit2genk}

The last case we consider is the generalization of the Limit 2 \eqref{eq:Limit1scaling} to higher $k$
\begin{equation}
    \textbf{Limit 2}:\qquad s_i=x^{2\phi_i}\,, \quad v_i=\mathrm{e}^{2\pi i\beta\nu_i}\,,\quad \xi,\phi_i,\nu_i\, \,\text{ finite}\,,\quad i=1,\cdots,k\,.
\end{equation}
The condition $\prod_{i=1}^k v_i=1$ translates into $\sum_{i=1}^k\nu_i=0$, so there are still only $k-1$ independent $\nu_i$.

As we did for $k=1$, on the SQED with $k$ flavors side we also take
\begin{equation}
    u=\mathrm{e}^{2\pi i\beta\sigma}\,.
\end{equation}
Using \eqref{eq:qPochhLimit}, we then get that in the $\beta\rightarrow 0$ limit
\begin{align}\label{eq:k_SQED_2d}
    \mathcal{I}_{\mathrm{A}}\xrightarrow[]{\beta \rightarrow0} \beta^{1-k+4\sum_i\phi_i}\sum_{m\in \mathbb{Z}}\int_{-\infty}^{\infty}\mathrm{d}\sigma (-1)^{km}\xi^m\prod_{i=1}^k\frac{\Gamma \left(\frac{1}{4}+\phi_i\pm \left(\frac{m}{2}+2\pi i (\sigma+\nu_i)\right)\right)}{\Gamma \left(\frac{3}{4}-\phi_i\pm \left(\frac{m}{2}-2\pi i (\sigma+\nu_i)\right)\right)}\,.
\end{align}

The mirror dual side is more complicated than the $k=1$ case as it is now a quiver gauge theory. Here we proceed similarly to what we did for Limit 1 on the SQED side. Namely, we introduce the new variables
\begin{equation}
    \label{eq:holomorphicgenk}
    Z_a=\mathrm{e}^{z_a}=u_ax^{-m_a}=u_a\mathrm{e}^{\frac{\beta m_a}{2}}\,,\quad \bar{Z}_a=\mathrm{e}^{\bar{z}_a}=u_a^{-1}x^{-m_a}=u_a^{-1}\mathrm{e}^{\frac{\beta m_a}{2}}\,,\qquad a=1,\cdots,k-1\,.
\end{equation}
For each $a$, we should then use a Dirac comb as in \eqref{eq:sumintegral} so to get $k-1$ complex integrals like \eqref{eq:measure}. Again, we can argue that only the zero sector of each sum coming from the Dirac combs contributes by studying the twisted superpotential that is generated in the $\beta\to0$ limit using \eqref{eq:qPochhLimit}. The final result is
\begin{equation}\label{eq:k_quiver_2d}
\begin{split}
    \mathcal{I}_{\mathrm{B}} \xrightarrow[]{\beta\to0} & \beta^{1-k+4\sum_i\phi_i}\prod_{i=1}^k\left[\frac{\Gamma(\frac{1}{2}+2\phi_i)}{\Gamma(\frac{1}{2}-2\phi_i)}\right]\\ &\times \prod_{a=1}^{k-1}\left[\int_{\mathbb{C}} \frac{\mathrm{d}^2Z_a}{\pi |Z_a|^2}\frac{|Z_a|^{-4\pi i (\nu_a-\nu_{a+1})+2(\phi_a-\phi_{a+1})}}{|1-\frac{Z_a}{Z_{a-1}}|^{1+4\phi_a}}\right]\frac{1}{|1-\frac{1}{Z_{k-1}\xi}|^{1+4\phi_k}}\,,
\end{split}
\end{equation}
where we have introduced $Z_0=1$ for convenience.

Equating \eqref{eq:k_SQED_2d} and \eqref{eq:k_quiver_2d} we find that the 3d index identity \eqref{eq:U1Indexk1} in the $\beta\to0$ Limit 2 gives the following finite identity:
\begin{align}\label{eq:Limit2genk}
    &\sum_{m\in \mathbb{Z}}\int_{-\infty}^{\infty}\mathrm{d}\sigma (-1)^{km}\xi^m\prod_{i=1}^k\frac{\Gamma \left(\frac{1}{4}+\phi_i\pm \left(\frac{m}{2}+2\pi i (\sigma+\nu_i)\right)\right)}{\Gamma \left(\frac{3}{4}-\phi_i\pm \left(\frac{m}{2}-2\pi i (\sigma+\nu_i)\right)\right)}\nn\\ 
    &\qquad=\prod_{i=1}^k\left[\frac{\Gamma(\frac{1}{2}+2\phi_i)}{\Gamma(\frac{1}{2}-2\phi_i)}\right] \prod_{a=1}^{k-1}\left[\int_{\mathbb{C}} \frac{\mathrm{d}^2Z_a}{\pi |Z_a|^2}\frac{|Z_a|^{-4\pi i (\nu_a-\nu_{a+1})+2(\phi_a-\phi_{a+1})}}{|1-\frac{Z_a}{Z_{a-1}}|^{1+4\phi_a}}\right]\frac{1}{|1-\frac{1}{Z_{k-1}\xi}|^{1+4\phi_k}}\,.
\end{align}
In Section \ref{sec:2dKS}, we will provide an explicit proof of this identity in Limit 2 using a two-dimensional version of the Kapustin--Strassler procedure which assumes only the Limit 1 identity \eqref{eq:Limit1k1} for $k=1$.

\subsubsection*{Physical Interpretation 1: Mirror symmetry}

The Limit 2 identity \eqref{eq:Limit2genk} has a direct interpretation in terms of a 2d mirror duality, which generalizes the one we had for $k=1$ in Section \ref{SubSection:Limit2k1}:
\begin{itemize}
    \item \textbf{Theory A:} GLSM with $U(1)$ gauge group and $k$ pairs of chiral fields $\Phi_i$, $\tilde{\Phi}_i$ with charges $\pm1$, where $i=1,\cdots,k$. There is no superpotential for the chirals.
    \item \textbf{Theory B:} LG model of a twisted chiral multiplet $\Sigma$, $k$ pairs of twisted chiral multiplets $Y^1_{i}$, $Y^2_{i}$ with $2\pi $-periodic imaginary part and a twisted superpotential
    \begin{equation}
    \widetilde{\mathcal{W}}=-\frac{1}{4\pi}\left[\Sigma\sum^{k}_{i=1}(Y^1_i-Y^2_i)+i\mu \sum_{i=1}^k(\mathrm{e}^{-Y^1_i}+\mathrm{e}^{-Y^2_i})\right]+M_\alpha^iY^\alpha_i\,.
    \end{equation}
\end{itemize}

As usual, the $S^2$ partition function of the GLSM takes the general form
\begin{align}\label{eq:GLSM_k}
    \mathcal{Z}_{\text{A}}=\pi^{2k}\sum_{m\in\mathbb{Z}}e^{im\theta}\int_{-\infty}^\infty \mathrm{d}\sigma \; \mathrm{e}^{-8\pi^2 i  \sigma \eta} \prod_{i=1}^k\prod_{\alpha=1,2}\frac{\Gamma (\frac{r^i_\alpha}{2}-2\pi i\tilde{m}^i_\alpha-\frac{mQ_\alpha}{2}-2\pi i \sigma Q_\alpha)}{\Gamma (1-\frac{r^i_\alpha}{2}+2\pi i\tilde{m}^i_\alpha-\frac{mQ_\alpha}{2}+2\pi i \sigma Q_\alpha)}\,.
\end{align}
This agrees with the l.h.s.~of \eqref{eq:Limit2genk} provided that we use the gauge charges $Q_1=1$, $Q_2=-1$ and we identify the parameters as
\begin{equation}\label{eq:paridentificationgenk}
    \mathrm{e}^{i\theta}=(-1)^k\xi\,,\quad \eta=0\,,\quad \tilde{m}_\alpha^i=i\frac{\phi_i}{2\pi}+(-1)^{\alpha-1}\nu_i\,,\quad r_\alpha^i=\frac{1}{2}\,,\quad i=1,\cdots,k\,,\quad \alpha=1,2\,.
\end{equation}

The partition function of the mirror dual LG model is a direct generalization of equation \eqref{eq:LGk1} as given in \cite{Gomis:2012wy}
\begin{align}\label{eq:LG_k}             
    \mathcal{Z}_{\text{B}}&=\sum_{m\in\mathbb{Z}}\mathrm{e}^{im\theta}\int_{-\infty}^\infty \mathrm{d}\sigma \mathrm{e}^{-8\pi^2 i  \sigma \eta}\prod_{i=1}^k\prod_{\alpha=1,2}\left[\int_{-\infty}^\infty \mathrm{d}x_i^\alpha\int_{-\pi}^{\pi}\mathrm{d}y_i^\alpha\right]\nn\\
    &\times\mathrm{e}^{4\pi i\sigma (x_i^1-x^2_i) +im (y_i^1-y^2_i)+2i\mathrm{e}^{-x_i^\alpha }\sin(y_i^\alpha)+(4\pi i \tilde{m}^i_\alpha-q^i_\alpha) x_i^\alpha}\,.
\end{align}

The strategy to relate the partition function of the LG model to the r.h.s.~of \eqref{eq:Limit2genk} should be by now clear. As in the previous cases, we perform the change of variables
\begin{equation}
    x^{\pm}_i=\frac{x^1_i\pm x^2_i}{2}\,,\qquad y^{\pm}_i=\frac{y^2_i\pm y^1_i}{2}\,,\qquad i=1,\cdots,k
\end{equation}
and we perform the integrals over $x^+_i$, $y^+_i$ using \eqref{eq:Bessel}
\begin{equation}
\begin{split}
    \mathcal{Z}_{\text{B}}=&\pi^k\prod_{a=1}^k\left[\frac{\Gamma(\frac{1}{2}+2\phi_a)}{\Gamma (\frac{1}{2}-2\phi_a)}\right]\sum_{m\in\mathbb{Z}}(-1)^{km}\xi^m \int_{-\infty}^\infty \mathrm{d}\sigma\prod_{a=1}^k\int_\mathbb{C} \mathrm{d}^2 z_a \frac{1}{|z_a|^{\frac{3}{2}-2\phi_a+4\pi i (\sigma+\nu_a)}}\left(\frac{z_a}{\bar{z}_a}\right)^{\frac{m}{2}}\frac{1}{|1-z_a|^{1+4\phi_a}}\,,
\end{split}
\end{equation}
where we have substituted the identifications \eqref{eq:paridentificationgenk} and defined $z_a=\mathrm{e}^{2(x^-_a-iy^-_a)+i\pi}$ as before. Now we can perform the sum-integral for the field strength multiplet $\Sigma$, yielding delta-functions
\begin{equation}
    \mathcal{Z}_{\text{B}}=\pi^{k+1}\prod_{a=1}^k\left[\frac{\Gamma(\frac{1}{2}+2\phi_a)}{\Gamma (\frac{1}{2}-2\phi_a)}\right]\prod_{a=1}^k\int_\mathbb{C} \mathrm{d}^2 z_a\frac{\delta(\prod_{a=1}^k|z_a|-1)\delta(\sum_{a=1}^k\arg(z_a)+\arg(\xi))}{\prod_{a=1}^k|z_a|^{\frac{3}{2}-2\phi_a+4\pi i \nu_a}|1-z_a|^{1+4\phi_a}}\,.
\end{equation}
The last step to get the r.h.s.~of \eqref{eq:Limit2genk} is to perform the change of variables
\begin{equation}
    Z_a=\prod_{i=1}^a z_i\,,\qquad a=1,\cdots,k
\end{equation}
and use the delta-function to remove the integral over $Z_k$. We have thus shown that the identity between the $S^2$ partition functions of the 2d mirror dual theories is equivalent to the Limit 2 identity in \eqref{eq:Limit2genk}.

\subsubsection*{Physical Interpretation 2: $A_{k-1}$ Toda CFT}

The r.h.s.~of \eqref{eq:Limit2genk} still has the form of a Coulomb gas integral. Unlike the previous cases, this is a multi-dimensional integral. However, there is no Vandermonde determinant structure relating the various integration variables, which instead appears in the Liouville Coulomb gas integrals \eqref{eq:LiouvilleCorr}. The correct interpretation is still in terms of a Coulomb gas integral, but in the 2d $A_{k-1}$ Toda CFT. 

Toda theory is a CFT of $k-1$ scalar fields $\vec{\phi}(z,\bar{z})$ valued in the Cartan of $A_{k-1}$ with an exponential interaction associated to the simple roots $\vec{e}_l$ with $l=1,\cdots,k-1$
\begin{equation}
    \mathcal{L}=\frac{1}{4\pi}(\partial_m\vec{\phi},\partial^m\vec{\phi)}+\mu\sum_{l=1}^{k-1}\mathrm{e}^{b(\vec{e}_l,\vec{\phi})} \,,
\end{equation}
where $\mu$ is the cosmological constant, $b$ the Toda coupling and $(\cdot,\cdot)$ is the standard inner product in the root space normalized such that $(\vec{e}_l,\vec{e}_l)=2$. The Liouville theory corresponds to the $A_1$ case. The background charge is
\begin{equation}
    \vec{Q}=Q\vec{\rho}\,,\qquad Q=b+b^{-1}\,,
\end{equation}
where $\vec{\rho}$ is the Weyl vector of $A_{k-1}$, which is half of the sum of all positive roots
\begin{equation}
    \vec{\rho}=\frac{1}{2}\sum_{\vec{e}\in\Delta_+}\vec{e}\,,\qquad \Delta_+=\left\{\vec{e}_l+\vec{e}_{l+1}+\cdots+\vec{e}_m\,|\,1\leq l\leq m\leq k-1\right\}\,.
\end{equation}
The central charge is
\begin{equation}
    c=(k-1)+12(\vec{Q},\vec{Q})=(k-1)(1+k(k+1)Q^2)
\end{equation}
and the basic objects are the vertex operators
\begin{equation}
    V_{\vec{\alpha}}(z,\bar{z})=\mathrm{e}^{(\vec{\alpha},\vec{\phi}(z,\bar{z}))}
\end{equation}
with dimensions
\begin{equation}
    \Delta(\vec{\alpha})=\frac{1}{2}(\vec{\alpha},2\vec{Q}-\vec{\alpha})\,.
\end{equation}

As discussed in \cite{Fateev:2007ab,Fateev:2008bm}, one can still find Coulomb gas integral expressions for correlation functions of vertex operators by integrating over the zero modes of the fields $\vec{\phi}(z,\bar{z})$, similarly to the Liouville case. For correlators of the form
\begin{equation}
    \langle V_{\vec{\alpha}_1}(y_1,\bar{y}_1)\dots V_{\vec{\alpha}_M}(y_M,\bar{y}_M)\rangle
\end{equation}
we can write a Coulomb gas integral provided that the following neutrality conditions holds:
\begin{equation}\label{eq:neutralityToda}
    \vec{\alpha}=\sum_{i=1}^M\vec{\alpha}_i=2\vec{Q}-b\sum_{l=1}^{k-1}N_l\vec{e}_l\,.
\end{equation}
One of the key results of \cite{Fateev:2007qn,Fateev:2007ab,Fateev:2008bm} is that in some cases the Coulomb gas integrals can be brought to a form where it is possible to perform an analytic continuation in the number of screening charges $N_l$, so to lift the neutrality condition and reconstruct the full correlator for generic values of the momenta.

In particular, some exact expressions for the three-point and four-point correlation functions of vertex operators have been worked out in \cite{Fateev:2007ab,Fateev:2008bm} when some of the operators are degenerate or demi-degenerate, i.e.~their momenta are aligned with some of $A_{k-1}$ weights. For example, for the three-point function of two generic and one semi-degenerate vertex operators we have
\begin{equation}\label{eq:toda_3pt}
\begin{split}
    &\langle V_{\vec{\alpha}_2}(\infty)  V_{\varkappa \vec{\omega}_{k-1}-b\vec{\omega}_1}(1)  V_{\vec{\alpha}_1}(0) \rangle =\\
    &=\left[\pi \mu \gamma(b^2) b^{2-2b^2}\right]^{\frac{(2\vec{Q}-\vec{\alpha},\vec{\rho})}{b}}\frac{(\Upsilon(b))^{k-1}\Upsilon(\varkappa)\prod_{\vec{e}\in\Delta_+}\Upsilon\left((\vec{Q}-\vec{\alpha}_1,\vec{e})\right)\Upsilon\left((\vec{Q}-\vec{\alpha}_2,\vec{e})\right)}{\prod_{l,m=1}^k \Upsilon\left(\frac{\varkappa+b}{k}+(\vec{\alpha}_1-\vec{Q},\vec{h}_l)+(\vec{\alpha}_2-\vec{Q},\vec{h}_l)-b\delta_{lm}\right)}\,,
\end{split}
\end{equation}
where $\vec{\omega}_1$ and $\vec{\omega}_{k-1}$ are the highest weights of the fundamental and of the antifundamental representation of $A_{k-1}$ respectively, while 
\begin{equation}
    \vec{h}_m=\omega_1-\sum_{l=1}^{m-1}\vec{e}_l
\end{equation}
are the remaining weights of the fundamental representation. As mentioned before, the way this expression was derived in \cite{Fateev:2007ab,Fateev:2008bm} consists of first writing the Coulomb gas integral for this three-point function for generic number of screening charges $N_l$ and then, using various integral identities, re-expressing it in an equivalent form that is suitable for analytic continuation to generic values of $N_l$. In this way, the neutrality condition is lifted and the correlator for generic values of $\vec{\alpha}_1$, $\vec{\alpha}_2$ and $\varkappa$ is reconstructed. The result \eqref{eq:toda_3pt} is a generalization of the DOZZ formula of the three-point function of Liouville, which was rederived using a similar strategy to the one we have just summarized in \cite{Fateev:2007qn}.

Of our interest is the following result for the four-point function of two generic, one semi-degenerate and one degenerate vertex operators:
\begin{equation}\label{eq:toda_4pt}
\begin{split}
    &\frac{\langle V_{\vec{\alpha}_2}(\infty)V_{\varkappa \vec{\omega}_{k-1}}(1)V_{-b\vec{\omega}_1}(z,\bar{z})V_{\vec{\alpha}_1}(0)  \rangle}{\langle V_{\vec{\alpha}_2}(\infty)  V_{\varkappa \vec{\omega}_{k-1}-b\vec{\omega}_1}(1)  V_{\vec{\alpha}_1}(0) \rangle} =\\
    &= |z|^{2b(\vec{\alpha}_1,\vec{h}_1)}|1-z|^{\frac{2b\varkappa}{k}} \int_{\mathbb{C}^{k-1}} \prod_{i=1}^{k-1}\mathrm{d}^2 t_i |t_i|^{2(A_i-B_i)}|t_i-t_{i+1}|^{2(B_i-A_{i+1}-1)}|t_1-z|^{-2A_1}\,,
\end{split}
\end{equation}
where for simplicity we have introduced the extra parameter $t_k=1$ and we have defined
\begin{align}\label{eq:AB}
    A_i&=\frac{b\varkappa}{k}-\frac{k-1}{k}b^2+b(\vec{\alpha}_1-\vec{Q},\vec{h}_1)+b(\vec{\alpha}_2-\vec{Q},\vec{h}_i)\,,\nn\\
    B_i&=1+b(\vec{\alpha}_1-\vec{Q},\vec{e}_1+\cdots+\vec{e}_i)\,.
\end{align}
Similarly to the three-point function, this expression is obtained starting from the Coulomb gas integral and rewriting it in a form where analytic continuation in the number of screening charges is possible. Hence, this result is valid for generic $\vec{\alpha}_1$, $\vec{\alpha}_2$ and $\varkappa$ with no need to impose any neutrality condition.

The r.h.s.~of \eqref{eq:Limit2genk} precisely takes the form of the integral in \eqref{eq:toda_4pt}
\begin{equation}
\begin{split}
    \int_{\mathbb{C}^{k-1}} \prod_{a=1}^{k-1}&\left[\frac{\mathrm{d}^2Z_a}{ \pi |Z_a|^2}\frac{|Z_a|^{-4\pi i (\nu_a-\nu_{a+1})+2(\phi_a-\phi_{a+1})}}{|1-\frac{Z_a}{Z_{a-1}}|^{1+4\phi_a}}\right]\frac{1}{|1-\frac{1}{Z_{k-1}\xi}|^{1+4\phi_k}}
    \\&=\frac{|1-z|^{-\frac{2b\varkappa}{k}}}{|z|^{2b(\vec{\alpha}_1,\vec{h}_1)}}\frac{\langle V_{\vec{\alpha}_2}(\infty)V_{\varkappa \vec{\omega}_{k-1}}(1)V_{-b\vec{\omega}_1}(z,\bar{z})V_{\vec{\alpha}_1}(0)  \rangle}{\langle V_{\vec{\alpha}_2}(\infty)  V_{\varkappa \vec{\omega}_{k-1}-b\vec{\omega}_1}(1)  V_{\vec{\alpha}_1}(0) \rangle}~,
\end{split}
\end{equation}
up to the change of variables $t_{k-a}=Z_{a}$ and with the identification of parameters
\begin{equation}
\begin{split}
    &z=\xi^{-1}\,,\\
    &A_i= \frac{1}{2}+\phi_k+\phi_{k-i+1}+2\pi i (\nu_{k-i+1}-\nu_k)\,, \qquad\ i=1,\cdots,k \,\\
    &B_i=1+\phi_k-\phi_{k-i}+2\pi i (\nu_{k-i}-\nu_k)\,, \qquad\ i=1,\cdots, k-1\,.
\end{split}
\end{equation}
These relations can be inverted and expressed in terms of the momenta using \eqref{eq:AB}
\begin{align}
    \phi_i&=\frac14+\frac12\left(A_{k-i+1}-B_{k-i}\right)\nn\\
    &=-\frac14+\frac b2\left[\frac{\varkappa}{k}-\frac{k-1}{k}b+\bigl(\vec{\alpha}_1+\vec{\alpha}_2-2\vec Q,\vec h_{k-i+1}\bigr)\right]\,, \qquad i=1,\cdots,k\,,\nn\\
    \nu_i&=\frac{1}{4\pi i}\left[A_{k-i+1}+B_{k-i}-\frac32-\frac1k\left(\sum_{j=1}^k A_j+\sum_{j=1}^{k-1}B_j-\frac{3k}{2}+1\right)\right]\nn\\
    &=\frac{b}{4\pi i}\bigl(\vec{\alpha}_2-\vec{\alpha}_1,\vec h_{k-i+1}\bigr)\,,\qquad i=1,\cdots,k\,,
\end{align}
where we introduced $B_0=1$ for convenience. To derive these relations, we also used that for $A_{k-1}$
\begin{equation}
    \sum_{l=1}^{k} h_l=0\,.
\end{equation}
Our result is compatible with the findings of \cite{Doroud:2012xw} that the $S^2$ partition function of SQED with $k$ flavors, which is the l.h.s.~of our Limit 2 identity \eqref{eq:Limit2genk}, coincides with the Toda four-point function of \eqref{eq:toda_4pt} for generic $\vec{\alpha}_1$, $\vec{\alpha}_2$ and $\varkappa$.

As for Limit 1, we see that the general dictionary between the parameters of the GLSM and those of the Toda correlator are
\begin{align}
    \text{FI}\,\,&\longleftrightarrow\,\,\text{position}\nn\\
    \text{masses}\,\,&\longleftrightarrow\,\,\text{momenta}\,.
\end{align}
In particular, on the GLSM side we have $2k-1$ mass parameters $\phi_i$, $\nu_i$ due to the condition $\sum_{i=1}^k\nu_i=0$ and accordingly on the CFT side we have $2k-1$ momenta encoded in $\vec{\alpha}_1$, $\vec{\alpha}_2$ and $\varkappa$.


\section{2d version of the Kapustin--Strassler derivation}
\label{sec:2dKS}

In this section we present a direct derivation of the generic $k$ identities obtained in Limit 1 \eqref{eq:Limit1k} and in Limit 2 \eqref{eq:Limit2genk}, which only relies on the Limit 1 identity for $k=1$ \eqref{eq:Limit1k1}. Hence, the latter is the only fundamental identity, from which all the others can be derived.

The derivation consists of a two-dimensional version of the Kapustin--Strassler derivation of 3d abelian mirror symmetry \cite{Kapustin:1999ha}. As we have reviewed in Section \ref{sec:3d}, this relies on the fact that the $k=1$ duality can be viewed as a functional Fourier transform, implemented by gauging a $U(1)$ symmetry. We have then seen in Section \ref{sec:SQED/XYZ} that a completely analogous interpretation is possible for the $k=1$ identities that we have found in the $\beta\to0$ limit of the 3d index identities, as well as the associated 2d mirror dualities related by gauging a $U(1)$ $(-1)$-form symmetry. Thus, it is not surprising that a 2d version of the Kapustin--Strassler piecewise derivation exists.

\subsection*{Limit 1 identity}

We begin with the derivation of the Limit 1 identity \eqref{eq:Limit1k}. Starting from its r.h.s.~we can replace the contribution of each of the $k$ pairs of chiral multiplets using \eqref{eq:Limit1k1}. This introduces $k$ auxiliary Coulomb gas-like integrals
\begin{align}
\begin{split}
   \text{r.h.s.~of \eqref{eq:Limit1k}} &=\prod_{i=1}^k\left[\frac{\Gamma(\frac{1}{2}+2\phi_i)}{\Gamma(\frac{1}{2}-2\phi_i)}\right] \sum_{\vec{m}\in\mathbb{Z}^{k-1}}\prod_{a=1}^{k-1}\left[\int_{-\infty}^{\infty} \mathrm{d}\sigma_a\left(\frac{v_a}{v_{a+1}}\right)^{-m_a}\right]\\
    &\times\prod_{i=1}^{k}\left[\frac{\Gamma(\frac{1}{4}-\phi_i\pm(\tau\delta_{i,k}+\frac{m_i-m_{i-1}}{2}+2\pi i (\sigma_{i-1}-\sigma_i)) }{\Gamma(\frac{3}{4}+\phi_i\pm(\tau\delta_{i,k}+\frac{m_{i-1}-m_{i}}{2}+2\pi i (\sigma_{i-1}-\sigma_i)) }\right]\\
    &=\sum_{\vec{m}\in\mathbb{Z}^{k-1}}\prod_{a=1}^{k-1}\left[\int_{-\infty}^{\infty} \mathrm{d}\sigma_a\left(\frac{v_a}{v_{a+1}}\right)^{-m_a}\right]\prod_{i=1}^k\left[\int _{\mathbb{C}}\frac{\mathrm{d}^2Z_i}{\pi}\right] \\
    &\times\prod_{i=1}^{k}\frac{1}{\left|Z_i\right|^{\frac{3}{2}+2\phi_i+2\tau\delta_{i,k}+4\pi i (\sigma_{i-1}-\sigma_i)}\left|1-Z_i\right|^{1-4\phi_i}}\left(\frac{Z_i}{\bar{Z}_i}\right)^{\frac{m_i-m_{i-1}}{2}}\,.
    \end{split}
\end{align}

Next, we observe that the original $k-1$ integrals over $\sigma_a$ yield delta functions while the $k-1$ sums over $m_a$ yield periodic delta functions, which enforce
\begin{align}
    &|Z_i|=|Z_j|\,,\qquad \text{for any }i,j\in\{1,\dots,k\}\,,\nn\\
    &\mathrm{arg}(Z_i v^{-1}_i)=\mathrm{arg}(Z_{i+1} v^{-1}_{i+1})\,,\qquad i=1,\cdots,k-1\,.
\end{align}
We can then use these delta functions to get rid of $k-1$ out of the $k$ auxiliary complex integrals that we introduced at the previous step, so to eventually land on the l.h.s.~of \eqref{eq:Limit1k}
\begin{align}
\begin{split}
   \text{r.h.s.~of \eqref{eq:Limit1k}}&=\frac{1}\pi\int_{\mathbb{C}} \mathrm{d}^2Z\frac{1}{|Z|^{2+2\tau}}\prod_{i=1}^k\left[\frac{1}{|v_iZ|^{-\frac{1}{2}+2\phi_i}|1-Zv_i|^{1-4\phi_i}}\right]\\
    &=\text{l.h.s.~of \eqref{eq:Limit1k}}\,.
    \end{split}
\end{align}

\subsection*{Limit 2 identity}

Let us now move to the derivation of the Limit 2 identity \eqref{eq:Limit2genk}. Starting from its l.h.s. we again trade the contribution of each of the $k$ pairs of chiral multiplets for $k$ auxiliary Coulomb gas-like integrals using \eqref{eq:Limit1k1}
\begin{align}\label{eq:Limit2KSproof}
    \text{l.h.s.~of \eqref{eq:Limit2genk}}&=\sum_{m\in \mathbb{Z}}\int_{-\infty}^{\infty}\mathrm{d}\sigma (-1)^{km}\xi^m\prod_{i=1}^k\frac{\Gamma \left(\frac{1}{4}+\phi_i\pm \left(\frac{m}{2}+2\pi i (\sigma+\nu_i)\right)\right)}{\Gamma \left(\frac{3}{4}-\phi_i\pm \left(\frac{m}{2}-2\pi i (\sigma+\nu_i)\right)\right)}\nn\\ 
    &= \sum_{m\in \mathbb{Z}}\int_{-\infty}^{\infty}\mathrm{d}\sigma \ \xi^{m}\prod_{i=1}^k\left[\frac{\Gamma(\frac{1}{2}+2\phi_i)}{\Gamma(\frac{1}{2}-2\phi_i)}\int_{\mathbb{C}} \frac{\mathrm{d}^2z_i}{\pi}\frac{1}{|z_i|^{\frac{3}{2}-2\phi_i+4\pi i (\sigma+\nu_i)}|1-z_i|^{1+4\phi_i}}\left(\frac{z_i}{\bar{z}_i}\right)^{\frac{m}{2}}\right]\,.
\end{align}

Next, we observe that the original integral over $\sigma$ and sum over $m$ yield respectively a delta and a periodic delta function, which enforce
\begin{equation}
    \prod_{i=1}^k|z_i|=1\,,\qquad \sum_{i=1}^k\mathrm{arg}(z_i)+\mathrm{arg}(\xi)=0\,.
\end{equation}
After performing the change of variables
\begin{equation}
    Z_a=\prod_{i=1}^a z_i\,,\qquad a=1,\cdots,k\,,
\end{equation}
we can then use these delta functions to get rid of one out of the $k$ auxiliary complex integrals that we introduced at the previous step, so to eventually land on the r.h.s.~of \eqref{eq:Limit2genk}
\begin{align}\label{eq:Limit2KSproof}
    \text{l.h.s.~of \eqref{eq:Limit2genk}}&=\prod_{i=1}^k\left[\frac{\Gamma(\frac{1}{2}+2\phi_i)}{\Gamma(\frac{1}{2}-2\phi_i)}\right] \prod_{a=1}^{k-1}\left[\int_{\mathbb{C}} \frac{\mathrm{d}^2Z_a}{\pi |Z_a|^2}\frac{|Z_a|^{-4\pi i (\nu_a-\nu_{a+1})+2(\phi_a-\phi_{a+1})}}{|1-\frac{Z_a}{Z_{a-1}}|^{1+4\phi_a}}\right]\frac{1}{|1-\frac{1}{Z_{k-1}\xi}|^{1+4\phi_k}}\nn\\
    &=\text{r.h.s.~of \eqref{eq:Limit2genk}}\,.
\end{align}


\section{2d/2d correspondence from $(A_{k-1},A_{N-1})$ Argyres--Douglas theory}
\label{sec:2d2dAD}

The integral identities from the previous sections find interesting application in the 2d/2d correspondence of \cite{Rastelli:2025nyn}. In this section we will discuss an example where we will in particular make use of the Limit 2 identity \eqref{eq:Limit2genk}.

The construction of \cite{Rastelli:2025nyn} is summarized in Figure \ref{fig:diagram2d2d}. The starting point is a 4d $\mathcal{N}=2$ SCFT $T$ on a suitable background with topology $S^2\times\Sigma_{g,n}$. On $\Sigma_{g,n}$ we perform a twist by the Cartan of the $SU(2)_R$ R-symmetry. This implies that 2d (2,2) supersymmetry is preserved in the two orthogonal directions. Such a 2d (2,2) theory obtained from twisted dimensional reduction of the 4d theory $T$ on $\Sigma_{g,n}$, denoted by $\FF[T;\Sigma_{g,n}]$,\footnote{We sometimes omit the additional data at the punctures that enter in the definition of the theory.} can be placed supersymmetrically on $S^2$ \cite{Benini:2012ui,Doroud:2012xw,Closset:2014pda,Gerchkovitz:2014gta}. In fact, there exist two ways to do this, which preserve respectively an $SU(2|1)_A$ and $SU(2|1)_B$ symmetry, where the first contains the $U(1)_V$ vector R-symmetry of (2,2) theories, while the second contains the $U(1)_A$ axial R-symmetry. For the construction of \cite{Rastelli:2025nyn}, one chooses the B-type $S^2_B$ background,\footnote{The A-type $S^2_A\times\Sigma_{g,n}$ is also important as it is related to the works of \cite{Nekrasov:2009rc,Nekrasov:2009ui,Nekrasov:2009uh,Nekrasov:2014xaa}. It leads to a 2d TQFT on $\Sigma_{g,n}$.} since this implies that instead reducing the 4d theory $T$ on $S^2_B$ leads to a 2d non-unitary CFT $\CC[T]$ whose chiral algebra is the same $\VV[T]$ that we associate to $T$ via the standard SCFT/VOA correspondence \cite{Beem:2013sza}. The reason is that the 4d and 2d R-symmetries are related by
\begin{equation}
    U(1)_V=U(1)_R\,,\qquad U(1)_A=U(1)_r\,,
\end{equation}
where $U(1)_R$ is the Cartan of $SU(2)_R$ that is preserved by the twist. Moreover, the $S^2_B$ background can be understood as gluing two hemispheres with opposite $U(1)_A=U(1)_r$ twist \cite{Gomis:2012wy}, and each of these two hemispheres realizes the chiral and anti-chiral algebra of $T$ via the construction of \cite{Oh:2019bgz,Jeong:2019pzg,Dedushenko:2023cvd}, which uses Kapustin's holomorphic-topological twist \cite{Kapustin:2006hi}.

In summary, we have
\begin{itemize}
    \item $SU(2)_R$ twist on $\Sigma_{g,n}$, leading to a 2d (2,2) theory $\FF[T;\Sigma_{g,n}]$ on $S^2_B$;
    \item the B-type $S^2_B$ background, leading to the 2d non-unitary CFT $\CC[T]$ with chiral algebra $\VV[T]$.
\end{itemize}
The 2d/2d correspondence relates the $S^2_B$ partition function of the (2,2) theory with the correlation function on $\Sigma_{g,n}$ of $\CC[T]$
\begin{equation}\label{eq:2d2did}
    \mathcal{Z}_{S^2_B}[\FF[T;\Sigma_{g,n};O_1,\cdots,O_n]]\sim\langle O_1\cdots O_n\rangle_{\Sigma_{g,n}}^{\CC[T]}\,.
\end{equation}
The punctures on the Riemann surface are interpreted as insertions of primaries $O_i$ in the non-unitary CFT. Hence, there are different types of punctures depending on the choice of primaries, which lead to different (2,2) theories on the other side. From the 4d perspective, the primaries insertions correspond to surface defects of $T$ that wrap the two-sphere and are localized at points on the Riemann surface.

One important entry in the dictionary of the correspondence comes from the different interpretation of the complex structure moduli of $\Sigma_{g,n}$ on the two sides:
\begin{equation}\label{eq:dictionary}
    \text{chiral exactly marginal defs.~of }\FF[T;\Sigma_{g,n}]\,\,\leftrightarrow\,\,\text{conformal cross ratios of }\CC[T]\text{ correlator}\,.
\end{equation}
Each side of \eqref{eq:2d2did} depends on such parameters. The $\sim$ is due to the fact that the partition function of the (2,2) theory suffers from counterterm ambiguities and is thus defined only up to a prefactor that is a holomorphically factorized function of the exactly marginal parameters \cite{Jockers:2012dk,Gomis:2012wy,Gerchkovitz:2014gta}.

\subsection*{GLSMs for $(k,k+N)$ $W_k$ minimal model correlators}

A class of examples that was considered in \cite{Rastelli:2025nyn} was that of the $T=(A_1,A_{2M})$ Argyres--Douglas theories \cite{Argyres:1995jj,Argyres:1995xn,Xie:2012hs}, whose associated non-unitary CFTs are the sequence of $(2,2M+3)$ Virasoro minimal models. Correlation functions on the sphere of primary operators in these theories can be expressed as Coulomb gas integrals. Hence, in the spirit of the analysis we did in the previous sections, it is not surprising that one can find GLSMs whose two-sphere partition functions match the minimal models correlators. Via the 2d/2d correspondence, these GLSMs furnish descriptions of the $\FF[(A_1,A_{2M});\Sigma_{0,n}]$ theories corresponding to compactifications of $(A_1,A_{2M})$ on an $n$-punctured sphere.

Our previous results allow us to generalize the analysis of \cite{Rastelli:2025nyn} to the larger class of Argyres--Douglas theories
\begin{equation}
    T=(A_{k-1}, A_{N-1})\,,\qquad \mathrm{gcd}(k,N)=1\,.
\end{equation}
The VOAs of these Argyres--Douglas theories were identified in \cite{Cordova:2015nma,Xie:2016evu} to be the $(k,k+N)$ $\mathcal{W}_k$ algebra. This is the chiral algebra of the $(k,k+N)$ $W_k$ minimal model \cite{Beltaos:2010ka}. Hence, by the construction of \cite{Rastelli:2025nyn} we expect that
\begin{equation}
    \CC[(A_{k-1}, A_{N-1})]=(k,k+N)\,\,\,W_k\,\,\,\text{minimal model}\,,\qquad \mathrm{gcd}(k,N)=1\,.
\end{equation}

The central charge of these minimal models is 
\begin{equation}
    c=(k-1)\left(1-\frac{(k+1)N^2}{k+N}\right)\,.
\end{equation}
The primaries in the spectrum of a generic minimal model are labeled by two representations, $(\Lambda_+,\Lambda_-)$, of $\widehat{\mathfrak{su}}(k)_{L^\pm}$ at levels $L^\pm$ such that
\begin{equation}\label{eq:constraint}
    \vec{\Lambda}_\pm=\sum_{i=1}^{k-1}\lambda^\pm_i \vec{\omega}_i\,,\qquad \lambda^\pm_i\geq0\,,\qquad \sum_{i=1}^{k-1} \lambda^\pm_i \leq L^\pm\,,
\end{equation}
where the last inequality ensures that the additional affine Dynkin label $\lambda_0^\pm=L^\pm-\sum_{i=1}^{k-1}\lambda_i^\pm$ is non-negative. For the minimal models at hand we have
\begin{equation}
    L^+ =k+N-k=N\,,\qquad L^-=k-k=0\,.
\end{equation}
Hence, the primaries are labelled by only one $SU(k)$ representation $\Lambda$, and we shall denote them by $\phi_\Lambda$. Their conformal weights are
\begin{equation}
    h_{\Lambda}=\bar{h}_{\Lambda}=\frac{k(\Lambda,\Lambda)-2N(\rho,\Lambda)}{2(k+N)}\,.
\end{equation}

The $(k,k+N)$ $W_k$ minimal models can be understood as a rational specialization of the $A_{k-1}$ Toda CFT with an imaginary coupling $b$ given by \cite{Chang:2011vka,Alkalaev:2014sma,Papadodimas:2011pf}
\begin{equation}
    b^2=-\frac{k}{k+N}\,.
\end{equation}
The primaries correspond to the Toda vertex operators in the following way:
\begin{equation}\label{eq:primaryvertex}
    \phi_{\Lambda} \leftrightarrow V_{-b\Lambda}\,.
\end{equation}

We will consider the following class of four-point functions on a sphere:
\begin{equation}\label{eq:minmodcorr}
    \langle \phi_{\vec{\Lambda}_2}(\infty)\phi_{s \vec{\omega}_{k-1}}(1)\phi_{\vec{\omega}_1}(z,\bar{z})\phi_{\vec{\Lambda}_1}(0)  \rangle
\end{equation}
where from \eqref{eq:constraint} we have
\begin{equation}
    \vec{\Lambda}_a=\sum_{i=1}^{k-1}\lambda_{a,i} \vec{\omega}_i\,,\qquad \lambda_{a,i}\geq0\,,\qquad \sum_{i=1}^{k-1} \lambda_{a,i} \leq N
\end{equation}
and
\begin{equation}
    s=1,\cdots,N\,.
\end{equation}
By the dictionary \eqref{eq:primaryvertex} we have that this correlator coincides with the Toda four-point function we studied in Section \ref{SubSection:Limit2genk}
\begin{equation}
    \langle V_{\vec{\alpha}_2}(\infty)V_{\varkappa \vec{\omega}_{k-1}}(1)V_{-b\vec{\omega}_1}(z,\bar{z})V_{\vec{\alpha}_1}(0)  \rangle\,,
\end{equation}
with the identification
\begin{equation}
    \vec{\alpha}_1=-b\Lambda_1\,,\qquad \vec{\alpha}_2=-b\Lambda_2\,,\qquad \varkappa=-bs\,.
\end{equation}
Hence, we can use the Limit 2 identity \eqref{eq:Limit2genk} to find a GLSM whose two-sphere partition function matches with this minimal model four-point function, up to a prefactor that is holomorphically factorized in $z$.

The GLSM is given by a $U(1)$ gauge group and $k$ pairs of chirals $\Phi_i$, $\tilde{\Phi}_i$ of charges $\pm1$, with $i=1,\cdots,k$. The parameters appearing in the partition function of this GLSM on the l.h.s.~of \eqref{eq:Limit2genk} are given in terms of the labels of the primaries as
\begin{align}\label{eq:phinu}
    \phi_i&=-\frac14+\frac{1}{2(k+N)}\left(s+k-1+N\left(k+1-2i\right)+k\sum_{r=k-i+1}^{k-1}(\lambda_{1,r}+\lambda_{2,r})-\sum_{r=1}^{k-1}r(\lambda_{1,r}+\lambda_{2,r})\right)\,,\nonumber\\
    \nu_i&=\frac{i}{4\pi(k+N)}\left[k\sum_{r=k-i+1}^{k-1}(\lambda_{1,r}-\lambda_{2,r})-\sum_{r=1}^{k-1}r(\lambda_{1,r}-\lambda_{2,r})\right]\,,\qquad i=1,\cdots,k \,.
\end{align}
Notice that $\phi_i$ are purely real, while $\nu_i$ are purely imaginary. This means that all real masses should be turned off. As we will see more explicitly soon in a subclass of examples, this is due to a superpotential among the chiral fields that breaks all the flavor symmetries, as expected since the original $(A_{k-1},A_{N-1})$ Argyres--Douglas theory does not possess a flavor symmetry if $\mathrm{gcd}(k,N)=1$. The values of $\phi_i$ and $\nu_i$ that we have found instead imply a shift of the R-charges of the chiral fields, which is again dictated by the superpotential. These R-charges are computed as (we use units in which the superpotential has R-charge 2)
\begin{equation}
    R_i=\frac{1}{2}+2\phi_i+4\pi i\nu_i\,,\qquad \tilde{R}_i=\frac{1}{2}+2\phi_i-4\pi i\nu_i\,,\qquad i=1,\cdots,k\,,
\end{equation}
where $R_i$ denotes the R-charge of $\Phi_i$ and $\tilde{R}_i$ that of $\tilde{\Phi}_i$. 

This GLSM can be understood via the 2d/2d correspondence as describing the compactification of the $(A_{k-1},A_{N-1})$ Argyres--Douglas SCFTs on a sphere with four punctures labelled by the choice of primaries in the correlator \eqref{eq:minmodcorr}, up to additional gauge singlet chiral fields as we will comment soon. It is important that all the fields of the GLSM are viewed as twisted fields, i.e.~twisted vector and twisted chiral fields. In this case, it is the B-type $S^2_B$ partition function that takes the form on the l.h.s.~of \eqref{eq:Limit2genk}. This also implies that the FI parameter $\xi$ is a chiral exactly marginal deformation, rather than twisted chiral as usual for a GLSM. Under the correspondence the FI is mapped to the position of the 4th operator in the CFT correlator, as expected from the general dictionary \eqref{eq:dictionary}. It descends from the single complex structure modulus of the four-punctured sphere.

For concreteness, let us consider for example the case in which
\begin{equation}\label{eq:family}
    N=k+1\,,\qquad \vec{\Lambda}_1=\vec{\Lambda}_2=\sum_{i=1}^{k-1}\vec{\omega}_i\,,\qquad s=2\,.
\end{equation}
Then \eqref{eq:phinu} reduces to
\begin{equation}
    \phi_i=\frac{2k+3-4i}{4(2k+1)}\,,\qquad \nu_i=0\,,\qquad i=1,\cdots,k\,.
\end{equation}
In particular the fact that all $\nu_i$ are turned off is due to $\Lambda_1=\Lambda_2$ and implies that the theory is non-chiral, that is $\Phi_i$ and $\tilde{\Phi}_i$ have the same R-charge
\begin{equation}
    R_i=\tilde{R}_i=\frac{2(k+1-i)}{2k+1}\,.
\end{equation}
These R-charges can be fixed by requiring the following superpotential to have R-charge 2:\footnote{More precisely, the superpotential constrains all the R-charges up to one parameter, which is however redundant since it can be re-absorbed with a gauge transformation.}
\begin{equation}
    \mathcal{W}=F^{2k+1}+F\Phi_1\widetilde{\Phi}_1+\sum_{i=1}^{k-1}\left[\left(\widetilde{\Phi}_{k+1-i}\right)^2\Phi_i\Phi_{i+1}+\left(\Phi_{k+1-i}\right)^2\widetilde{\Phi}_i\widetilde{\Phi}_{i+1}\right]\,,
\end{equation}
where $F$ is a gauge singlet chiral field of R-charge $\tfrac{2}{2k+1}$ that we need to add to the GLSM. 

As pointed out in \cite{Rastelli:2025nyn}, singlet fields just provide an overall prefactor in the two-sphere partition function and are thus part of the holomorphically factorized ambiguity of \eqref{eq:2d2did}. Moreover, the singlets are needed not only to write the superpotential, but also for the GLSM to reproduce some properties that are expected for the compactification of the 4d $\mathcal{N}=2$ SCFT, such as their central charges and elliptic genera. In fact, the GLSM that we have found in the family of examples \eqref{eq:family} reduces for $k=2$ to the one found in \cite{Rastelli:2025nyn} for the four-point function of the $\phi_{(2,1)}$ of the Lee--Yang minimal model, which is associated to the $(A_1,A_2)$ SCFT, up to a second singlet that couples to $\Phi_2\tilde{\Phi}_2$. It would be interesting to perform the same tests done in \cite{Rastelli:2025nyn} for the cases considered here and to work out all the singlets that are required. 

\acknowledgments

The work of SC is supported in part by NSF grant PHY-2513893 and by the Simons Foundation grant 681267 (Simons Investigator Award).

\appendix

\section{Supersymmetric partition functions}
\label{app:SUSYpf}

\subsection*{3d supersymmetric index}

In this appendix we summarize our conventions for the supersymmetric partition functions that we used in the main text. 

The first one is the supersymmetric index for three-dimensional $\mathcal{N}=2$ theories \cite{Bhattacharya:2008zy,Kim:2009wb,Imamura:2011su,Kapustin:2011jm,Dimofte:2011py}. To define it, we start from the superconformal index
\begin{equation}
    \mathcal{I}(x)=\mathrm{Tr}(-1)^{2J_3}x^{R+2J_3}\,.
\end{equation}
where $R$ is the generator of the $U(1)_R$ R-symmetry, while $J_3$ is the spin. The trace is taken over the states of the theory in radial quantization, that are annihilated by a supercharge $\mathcal{Q}$ such that
\begin{equation}
    \delta=\left\{\mathcal{Q},\mathcal{Q}^\dagger\right\}=E-J_3-R\,,
\end{equation}
so only the states with $\delta=0$ contribute to the index.

The supersymmetric index is a variant of the superconformal index, where the R-symmetry is not necessarily the superconformal one. It can also be viewed as a partition function on $S^2\times S^1$. From either perspective, it can be expressed for a Lagrangian theory with gauge group $G$ as
\begin{equation}
    \mathcal{I}(x)=\frac{1}{|W|}\sum_{\vec{m}\in \Gamma^\vee _G}\prod_{a=1}^{\mathrm{rk(G)}}\left[\oint \frac{du_a}{2\pi i u_a}\right] Z_{\mathrm{CS}}Z_{\mathrm{vec}}Z_{\mathrm{chir}} \,.
\end{equation}
where $|W|$ is the order of the Weyl group of $G$, the sum is over the magnetic fluxes $\vec{m}$ through $S^2$ valued in the co-weight lattice $\Gamma^\vee_G$ of the gauge group $G$, and the integral is over gauge holonomies $u_a=\mathrm{e}^{i\theta_a}$, $\theta_a\in[-\pi,\pi)$ wrapping the $S^1$ direction and valued in the maximal torus of $G$. 

The various factors contributing to the integrand are:
\begin{enumerate}
    \item The chiral multiplet contribution
    \begin{equation}
    Z_{\mathrm{chir}}:=\prod_{\vec{\rho}\in\Lambda}\frac{(\vec{u}^{-\vec{\rho}} x^{2-\vec{\rho}(\vec{m})-r}f^{-1};x^2)_\infty}{(\vec{u}^{\vec{\rho}} x^{-\vec{\rho}(\vec{m})+r}f;x^2)_\infty}\,,
\end{equation}
where $\vec{\rho}$ are the weights of the representation $\Lambda$ under which the chiral transforms, $r$ is its R-charge, and we introduced a fugacity $f$ for a possible $U(1)$ flavor symmetry (which can be one of the Cartans of a larger non-abelian flavor symmetry if the theory has multiple identical chirals).\footnote{One can also refine the index by a background flux for the flavor symmetry. This would appear in the same way as the gauge flux $m$ in the integrand, however we do not sum over it in the full index.} Here
\begin{equation}
    (z;q)_\infty=\prod_{k=0}^{\infty}(1-zq^k)\,,
\end{equation}
is the $q$-Pochhammer symbol.
    
\item  The vector multiplet contribution
    \begin{equation}
    Z_{\mathrm{vec}}=\prod_{\vec{\alpha}\in \Delta_+}x^{\vec{\alpha}(\vec{m})}(1-x^{-\vec{\alpha}(\vec{m})}\vec{u}^{\pm\vec{\alpha}})\,,
\end{equation}
where $\Delta_+$ is the set of positive roots of $G$.
\item The Chern--Simons contribution
\begin{equation}
    Z_{\mathrm{CS}}=\left[\prod_{a,b=1}^{\mathrm{rk}(G)}(-1)^{k_{ab}m_am_b}u_a^{k_{ab}m_b}\right]\left[\prod_{a=1}^{\mathrm{rk}(G)}u_a^nx^{2k_{R,a}m_a}\xi^{m_a}\right]\,,
\end{equation}
where we introduced a flux $n$ and a fugacity $\xi$ for a possible $U(1)$ topological symmetry. We also defined the effective gauge/gauge and gauge/R-symmetry CS couplings
\begin{equation}
    k_{ab}:=k^\mathrm{bare}_{ab}+\frac{1}{2}\sum_{i:\text{ chir}}\sum_{\vec{\rho}\in\Lambda_i}\rho_a\rho_b\,,\quad k_{R,a}:=k^\mathrm{bare}_{R,a}+\frac{1}{2}\sum_{i:\text{ chir}}\sum_{\vec{\rho}\in\Lambda_i} \left(1-r_i\right)\rho _a \,,
\end{equation}
with the index $i$ running over all the chirals of the theory, with corresponding representations $\Lambda_i$ and R-charges $r_i$.
\end{enumerate}
In the above expressions we have made use of some shorthand notations, such as
\begin{equation*}
    \vec{u}^{\vec{\rho}}=\prod_{a=1}^{\mathrm{rk}(G)}u_a^{\rho_a}\,,\quad \vec{\rho}(\vec{m})=\sum_{a=1}^\mathrm{rk(G)} \rho_a m_a
\end{equation*}

While the one above is the form we will find most convenient to work with, we note briefly that there exists another presentation of these quantities that is common in the literature:
\begin{enumerate}
    \item The chiral multiplet contribution
    \begin{equation}
    \tilde{Z}_{\mathrm{Chiral}}=\prod_{\vec{\rho}\in\Lambda}\frac{(\vec{u}^{-\vec{\rho}} x^{2+|\vec{\rho}(\vec{m})|-r}f^{-1};x^2)_\infty}{(\vec{u}^{\vec{\rho}} x^{|\vec{\rho}(\vec{m})|+r}f;x^2)_\infty}\,.
\end{equation}
    
\item  The vector multiplet contribution
    \begin{equation}
    \tilde{Z}_{\mathrm{vec}}:=\prod_{\vec{\alpha}\in \Delta_+}(1-x^{-|\vec{\alpha}(\vec{m})|}\vec{u}^{\pm\vec{\alpha}})\,.
\end{equation}
\item The Chern--Simons contribution
\begin{equation}
    Z_{\mathrm{CS}}=\left[\prod_{a=1}^{\mathrm{rk}(G)}(-1)^{Q_{a}(\vec{m})m_a}u_a^{Q_a(\vec{m})}\right]\left[\prod_{a=1}^{\mathrm{rk}(G)}u_a^n\xi^{m_a}\right]x^{2R(\vec{m})}\,,
\end{equation}
where we have defined the gauge and R-charges of a monopole of magnetic flux $\vec{m}$ as
\begin{equation}
\begin{split}
    &Q_{a}(\vec{m}):=k^{\mathrm{bare}}_{ab}m_b-\frac{1}{2}\sum_{i:\text{ chir}}\sum_{\rho\in\Lambda
    _i}\rho_a|\vec{\rho}(\vec{m})|\,,\\
    &R(\vec{m}):=k^\mathrm{bare}_{R,a}m_a+\frac{1}{2}\sum_{i:\text{ chir}}\sum_{\rho\in \Lambda_i}(1-r_i)|\vec{\rho}(\vec{m})|-\sum_{\vec{\alpha}\in \Delta_+} |\vec{\alpha}(\vec{m})|\,.
    \end{split}
\end{equation}
\end{enumerate}
The two formulations can be related to one another by noting that
\begin{equation}
    (-x^{-1}u)^{-\frac{|m|}{2}}\frac{(x^{2+|m|}u^{-1};x^2)_\infty}{(x^{|m|}u;x^2)_\infty}=(-x^{-1}u)^{\frac{m}{2}}\frac{(x^{2-m}u^{-1};x^2)_\infty}{(x^{-m}u;x^2)_\infty}\,.
\end{equation}

\subsection*{$S^2$ partition function}

The second class of partition functions that we considered is that of 2d $\mathcal{N}=(2,2)$ on a two-sphere. We only focus here on the A-type $S^2$ background and distinguish between the cases of a GLSM of vector and chiral fields \cite{Benini:2012ui,Doroud:2012xw}, and of a LG model of twisted chiral fields \cite{Gomis:2012wy}.

The partition function of a GLSM can be localized onto either Coulomb or Higgs branch configurations. For our pruposes we are only interested in the Coulomb branch localization, which gives\footnote{We introduce a normalization by some $\pi$ factor which is needed in order for the partition function of a GLSM to match with that of its LG mirror dual.}
\begin{equation}
    \mathcal{Z}=\frac{1}{|W|}\sum_{\vec{m}\in \Gamma^\vee _G}\prod_{a=1}^{\mathrm{rk(G)}}\left[\pi^2\int_{-\infty}^{\infty} \mathrm{d}\sigma_a\right]\mathcal{Z}_{\mathrm{cl}}\mathcal{Z}_{\mathrm{chi}}\mathcal{Z}_{\mathrm{vec}}\,.
\end{equation}
Similarly to the 3d index, the sum is over magnetic fluxes $\vec{m}$ through $S^2$ and the integral is over the Cartan subalgebra of the gauge group $G$, however this time the latter takes value on the real axis. 

The various contributions to the integrand are:
\begin{enumerate}
    \item The chiral contribution
    \begin{equation}
    \mathcal{Z}_{\mathrm{chir}}=\prod_{\vec{\rho}\in\Lambda} \frac{\Gamma \left(\frac{r-\vec{\rho}(\vec{m})}{2}-2\pi i(\vec{\rho}(\vec{\sigma}) +\tilde{m})\right)}{\Gamma \left(1-\frac{r+\vec{\rho}(\vec{m})}{2}+2\pi i(\vec{\rho}(\vec{\sigma}) +\tilde{m})\right)}\,,
\end{equation}
where we introduced a twisted mass $\tilde{m}$ which is the bottom component for a background field strength multiplet for a possible $U(1)$ flavor symmetry under which the chiral transforms. This can be repackaged with the R-charge $r$ into the complex combination
\begin{equation}
    M=\tilde{m}+i\frac{r}{2\pi}\,.
\end{equation}
\item The vector contribution
\begin{equation}
    \mathcal{Z}_{\mathrm{vec}}= \prod_{\vec{\alpha}\in \Delta_+}\left[\left(\frac{\vec{\alpha}(\vec{m})}{2}\right)^2+(\vec{\alpha}(\vec{\sigma}))^2\right]\,.
\end{equation}
This trivializes when the gauge group is Abelian, which is the only case we consider in this work.
\item The classical contribution
\begin{equation}
    \mathcal{Z}_{\mathrm{cl}}=\mathrm{e}^{i\theta \mathrm{Tr}(\vec{m})-8\pi^2 i \eta \mathrm{Tr}(\vec{\sigma})}\,,
\end{equation}
where $\eta$ is the FI parameter and $\theta$ is the theta angle. Here we are assuming that $G$ contains only one $U(1)$ factor, e.g.~for $G=U(N)$ we have $\mathrm{Tr}(\vec{m})=\sum_{a=1}^Nm_a$ and $\mathrm{Tr}(\vec{\sigma})=\sum_{a=1}^N\sigma_a$. In case of multiple $U(1)$s, we can introduce an FI parameter and a theta angle for each of them, with a straightforward generalization of the previous expression.
\end{enumerate}

For a 2d $\mathcal{N}=(2,2)$ LG model of $n$ twisted chiral fields and with twisted superpotential $\widetilde{\mathcal{W}}$, the $S^2$ partition function takes the form
\begin{equation}
    \mathcal{Z}=\left[\prod_{i=1}^n\int\mathrm{d}^2z_i\right]\mathrm{e}^{-4\pi i\left(\widetilde{\mathcal{W}}(z_i)+\overline{\widetilde{\mathcal{W}}}(\bar{z}_i)\right)}\,.
\end{equation}
For standard twisted chiral multiplets, the integration domain is $\mathbb{C}$. However, those appearing in the LG models that are mirror dual to a GLSM have a $2\pi$ periodic imaginary part
\begin{equation}
    z_i=x_i+iy_i\,,\qquad x_i\in\mathbb{R}\,,\qquad y_i\in [-\pi,\pi)\,.
\end{equation}

\section{Derivation of $\eqref{eq:ModifiedDFIntegrals}$}
\label{app:proofDF}

In this appendix we provide a derivation of \eqref{eq:ModifiedDFIntegrals}, which generalizes the Limit 1 identity for $k=1$ \eqref{eq:Limit1k1}. Our derivation is a small variant of \cite{Dotsenko:1984ad} (see also \cite{Kapec:2020xaj} for a review). We start by writing
\begin{align}\label{eq:Fint}
    F(a,b,c,d)&=\frac{1}{\pi}\int_\mathbb{C} \mathrm{d}^2z\, z^a\bar{z}^b(1-z)^c(1-\bar{z})^d\nn\\
    &=\frac{1}{\pi}\int_{-\infty}^{\infty} \mathrm{d}x\int_{-\infty}^{\infty} \mathrm{d}t \,(x+it)^a(x-it)^b(1-x-it)^c(1-x+it)^d\,,
\end{align}
where we defined
\begin{equation}
    z=x+it\,,\qquad \bar{z}=x-it\,.
\end{equation}
The idea now is to Wick rotate $t$ to 
\begin{equation}
    t=i\tau \mathrm{e}^{-i\epsilon}\,,\qquad \epsilon>0\,.
\end{equation}
However, we must check that 
\begin{enumerate}[label=\alph*)]
    \item we do not encounter any branch cuts for any $t,x\in\mathbb{R}$;
    \item the asymptotics are such the contribution from the arcs at infinity swept by the Wick rotation are vanishing.
\end{enumerate}

Let us check a) first. As $t\to0$, the integrand may lie on a branch-cut and thus be ill-defined. In order to avoid so, for the branch point at $x+it=0$, we require
\begin{equation}
  \lim_{\mu\to0}  (x+i\mu)^a(x-i\mu)^{b}-(x-i\mu)^a(x+i\mu)^{b}=0\,,\qquad x<0\,.
\end{equation}
This gives us the condition
\begin{equation}
    a-b\in\mathbb{Z}\,.
\end{equation}
Similarly, for the branch point at $x+it=1$, we require
\begin{equation}
  \lim_{\mu\to0}  (1-x-i\mu)^c(1-x+i\mu)^{d}-(1-x+i\mu)^c(1-x-i\mu)^{d}=0\,,\qquad x>1\,.
\end{equation}
This gives us the condition
\begin{equation}
    c-d\in\mathbb{Z}\,.
\end{equation}

Let us now turn to the asymptotic analysis of the contribution from the arcs at infinity in point b). We write $t=R\mathrm{e}^{i\theta}$ for $\theta\in\left[0,\frac{\pi}{2}-\epsilon\right]$, and study the integrand in the large $R$ asymptotics to find,
\begin{equation}
    \,(x+it)^a(x-it)^b(1-x-it)^c(1-x+it)^d\underset{R\to\infty}{\sim}R^{a+b+c+d}\,.
\end{equation}
Thus, in order for the regions at infinity not to contribute, we require
\begin{equation}
    \mathrm{Re}(a+b+c+d)<-1\,.
\end{equation}

Now that we have discussed the validity of Wick rotating, we proceed in an identical manner to \cite{Dotsenko:1984ad}, which we review here for convenience. Wick rotating turns the integrand to,
\begin{equation}
    (x-\tau+i\epsilon\tau)^a (x+\tau-i\epsilon\tau)^b (1-x+\tau-i\epsilon\tau)^c(1-x-\tau+i\epsilon\tau)^d\,.
\end{equation}
Switching to lightcone coordinates
\begin{equation}
    u=x-\tau\,,\quad v=x+\tau\,,
\end{equation}
gives the integrand (where we have rescaled $\epsilon\to\epsilon/2$),
\begin{equation}
    (u+i\epsilon(v-u))^a (v-i\epsilon(v-u))^b (1-u-i\epsilon(v-u))^c(1-v+i\epsilon(v-u))^d\,.
\end{equation}
Therefore, apart from the $\epsilon$ factors, our integrand factorizes into a product over advanced and retarded lightcone coordinates contributions. 

To see how the $\epsilon$ factors affect our integrands, let us figure out the positions of the poles and branch points in the complex $u$-plane in terms of $v$.
We have the branch points
\begin{equation}
    u_{0,<}=-\frac{i\epsilon v}{1-i\epsilon}\, \quad\mathrm{and}\quad u_{1,>}=\frac{1-i\epsilon v}{1-i\epsilon}\,.
\end{equation}
We will see that for different regions in $v\in\mathbb{R}$, the $u$-integral vanishes due to the dependence of the $\epsilon$ prescription on $v$. For $v>1$, the branch cut with branch point $u_{0,<}$ extends from $u=-i\epsilon|v|+O(\epsilon^2)$ leftwards (as indicated by the $<$ subscript), while the branch cut with branch point $u_{1,>}$ extends from $u=-i(\epsilon|v|-1)+O(\epsilon^2)$ rightwards (as indicated by the $>$ subscript). Thus, for $v>1$ both branch cuts are below the real $u$-axis. Since $a,c$ are chosen to have appropriate fall-off conditions
\begin{equation}
\mathrm{Re}(a+c)<-1\,,
\end{equation}
we could thus deform our contour in the upper-half plane freely, and thus we can see that the contribution of the $v>1$ region of the $u$ integral vanishes. A mirrored argument occurs for $v<0$. We are therefore left with $v\in[0,1]$, for which the branch cuts trap the original contour along the real axis in the $u$-plane. Condensing our integral onto one of the branches thus gives
\begin{equation}
  F(a,b,c,d)=-\frac{\sin{(\pi c)}}{\pi}  \int_0^1 \mathrm{d}v\, (v+i\epsilon)^b(1-v-i\epsilon)^d\int_1^\infty \mathrm{d}u \, u^a(u-1-i\epsilon)^c\,.
\end{equation}
In the above, the $\sin{(\pi c)}$ factor comes from the branch cut discontinuity when condensing the $u$-integral along the branch cut associated with $u_{1,>}$, and absorbs a factor of $2i$ from the Wick rotation and Jacobian associated with the lightcone coordinate change of variables. 

The standard Euler integral representations of the beta function
\begin{equation}
   B(a,c-a)=\frac{\Gamma(a)\Gamma(c-a)}{\Gamma(c)}= \int_0^1 dt \,t^{a-1}t^{c-a-1}\,,
\end{equation}
the reflection equation for $\Gamma$ functions
\begin{equation}
    \Gamma(1+c)=\frac{-\pi}{\sin(\pi c)\Gamma(-c)}\,,
\end{equation}
together with a change of variables $p=u^{-1}$, gives
\begin{align}
\begin{split}
    F(a,b,c,d)&=-\frac{\sin{(\pi c)}}{\pi}  \int_0^1 \mathrm{d}v\, (v)^b(1-v)^d\int_0^1 \mathrm{d}p \, p^{-a-c-2}(1-p)^c\,,\\
    &=-\frac{\sin{(\pi c)}}{\pi} B(1+b,1+d)B(-1-a-c,1+c)\,,\\
    &=-\frac{\sin{(\pi c)}}{\pi} \frac{\Gamma(b+1)\Gamma(d+1)}{\Gamma(2+b+d)}\frac{\Gamma(1+c)\Gamma(-(c+a+1))}{\Gamma(-a)}\,,\\
     &=\frac{\Gamma(b+1)\Gamma(d+1)}{\Gamma(2+b+d)}\frac{\Gamma(-(c+a+1))}{\Gamma(-a)\Gamma(-c)}\,.
\end{split}    
\end{align}
Thus completing our derivation of \eqref{eq:ModifiedDFIntegrals}. 

We point out that in our derivation we needed to assume
\begin{equation}
    \text{Re}\left(b,d\right)>-1\,,\text{ Re}\left(a+c,a+b+c+d\right)<-1\, \text{ and }\,a-b\,,c-d\in\mathbb{Z}\,,
\end{equation}
where to the conditions coming from the requirements a) and b) above we have also added those for the convergence of the integral after Wick rotation. We can however use the evaluation of the integral in terms of Gamma functions that we have derived in this range to analytically continue to more general values of $a$, $b$, $c$, $d$ as long as none of the arguments of the Gamma functions becomes a negative integer.

\section{A proof of the Limit 2 identity for $k=1$}
\label{app:Limit2k1residueproof}

In this appendix, we present a proof of the Limit 2 identity for $k=1$ in \eqref{eq:Limit2k1} based on a residue computation. Part of this computation has also appeared in \cite{Benini:2012ui}. Anticipating the generalization to generic values of $k$, we recognize that the integral on the l.h.s.~of \eqref{eq:Limit2k1} is a special type of Barnes integral, known as the Meijer G-function
\begin{align}
    G^{m,n}_{p,q}(\vec{a};\vec{b}|z)=\frac{1}{2\pi i}\int_{-i\infty}^{i\infty} \mathrm{d}s \,z^s \frac{\prod_{i=1}^m \Gamma(1-a_i+s)\prod_{j=1}^n \Gamma(b_j-s)}{\prod_{i=m+1}^p \Gamma(a_i-s)\prod_{j=n+1}^q \Gamma(1-b_j+s)}\,,
\end{align}
where $m,n,p,q\in \mathbb{Z}$, $0\leq m \leq p$, $0\leq n \leq q$ and $\vec{a}$ ($\vec{b}$) are $p$($q$)-dimensional parameters. We specialize to $z=1,s=2\pi i\sigma$. 

For the proper definition of $G^{m,n}_{p,q}$, the parameters $\vec{a}$, $\vec{b}$ are such that all the poles of $\Gamma(b_j-s)$ at $b_{j=1,\ldots,n}$ are on the right half-plane of the complex $s$-plane, and all the poles of $\Gamma(1-a_i+s)$ at $a_{i=1,\ldots,m}$ are on the left. Then, Slater's theorem relates Meijer G-function with the generalized hypergeometric function, by closing the integration contour in such a way that all the poles to the right are encircled. We are interested in the case where $m=n=k$, $p=q=2k$
\begin{align}\label{eq:Barnes_integral_Slater's}
    G^{k,k}_{2k,2k}(\vec{a};\vec{b}|1)=&\sum_{i=1}^k\frac{\prod_{j\neq i}^k \Gamma (b_j-b_i)\prod_{j=1}^k \Gamma (1+b_i-a_j)}{\prod_{j=k+1}^{2k} \Gamma (1+b_i-b_j)\Gamma (a_j-b_i)}\\
    &\times _{2k}F_{2k-1}\left(\{1+b_i-a_l\}_{l=1}^{2k};\{1+b_i-b_l\}_{l\neq i}^{2k};1\right).
\end{align}

Specializing to $k=1$, we get
\begin{equation}
\begin{split}
    \text{l.h.s.~of \eqref{eq:Limit2k1}}=\sum_{m\in \mathbb{Z}}(-\xi)^{m} G^{1,1}_{2,2}\left(\{1-\gamma+\frac{m}{2},1-\gamma-\frac{m}{2}\};\{\gamma+\frac{m}{2},\gamma-\frac{m}{2}\}|1\right)
\end{split}
\end{equation}
where $\gamma=\frac{1}{4}+\phi$. Using Slater's theorem \eqref{eq:Barnes_integral_Slater's} and the identity
\begin{equation}
    {}_2F_1\left(a,b;c;1\right) = \frac{\Gamma\left(c\right)\Gamma\left(c-a-b\right)}{\Gamma\left(c-a\right) \Gamma\left(c-b\right)}\,,\qquad \mathrm{Re}\left(c-a-b\right)>0\,,
\end{equation}
we get
\begin{align}
    \text{l.h.s.~of \eqref{eq:Limit2k1}}=\sum_{m\in \mathbb{Z}}(-\xi)^{m}\frac{\Gamma(2\gamma)\Gamma(1-4\gamma)}{\Gamma(1-2\gamma-m)\Gamma(1-2\gamma+m)\Gamma(1-2\gamma)}
\end{align}
for $4\gamma<1$. This can be matched precisely with the r.h.s.~of \eqref{eq:Limit2k1} by recognizing that
\begin{align}
    \sum_{m\in \mathbb{Z}}(-\xi)^{m}\frac{\Gamma(1-4\gamma)}{\Gamma(1-2\gamma-m)\Gamma(1-2\gamma+m)}=(1-\xi)^{-2\gamma}(1-\xi^{-1})^{-2\gamma}.
\end{align}


\bibliographystyle{JHEP}
\bibliography{refs}

\end{document}